\documentclass[aps,prd,preprint,groupedaddress]{revtex4}

\usepackage{epsfig}
\usepackage{amsmath}
\usepackage{slashbox}
\begin{document}

\def\beq{\begin{equation}}
\def\eeq{\end{equation}}
\def\bea{\begin{eqnarray}}
\def\eea{\end{eqnarray}}

\newcommand\ds{\displaystyle}

\title{Fragmentation Functions for Light Charged Hadrons with Complete
Quark Flavour Separation}
\author{S. Albino, B.A. Kniehl and G. Kramer}
\affiliation{{II.} Institut f\"ur Theoretische Physik, Universit\"at Hamburg,
             Luruper Chaussee 149, D-22761 Hamburg, Germany}
\date{\today}
\begin{abstract}
We present new sets of next-to-leading order fragmentation functions
describing the production of charged pions, kaons and protons from the gluon
and from each of the quarks, obtained by fitting to all relevant
data sets from $e^+ e^-$ annihilation. The individual light quark flavour fragmentation
functions are obtained phenomenologically for the first time by
including in the data the light quark tagging probabilities
obtained by the OPAL Collaboration.
\end{abstract}
\maketitle

\section{Introduction}

Theoretical predictions for future experiments are necessary for
determining the kinematic regions of validity of the Standard Model (SM). Such
predictions depend on constants which must be determined from past
experiments since these quantities are otherwise uncalculable, either
because no theory exists which can determine them from more
fundamental parameters, or because the solutions of the current theory
are insufficient to determine them from the SM parameters.

Quantum Chromodynamics (QCD), the theory of the strong interaction and
one of the theories that make up the SM, is required in the description of
processes involving hadrons. The best tool for solving QCD to perform such descriptions is
perturbation theory. However, perturbative QCD (pQCD) can only
describe the high energy components of the cross section, while a
process will contain low energy components if a hadron is in the
initial state or is observed in the final state. Fortunately, from the
Factorization Theorem, the low and high energy scale components of
such processes can be separated. The low energy components are
universal and so can be used to make predictions.  Since they cannot
yet be reliably calculated from QCD, they must be extracted from
experimental data.

The pQCD description of data involving the inclusive production of
hadrons requires fragmentation functions (FFs), which form the low
energy components of such processes and describe the inclusive
emission of a hadron from a quark or gluon (parton) for every momentum
fraction. One reason FFs are important is that
model independent predictions of LHC cross sections in which
a hadron is detected in the final state depend on them.
There are many theoretical obstacles to the extraction of
FFs from data: The Dokshitzer-Gribov-Lipatov-Altarelli-Parisi (DGLAP) \cite{dglap} 
evolution equation for FFs is only known
to next-to-leading order (NLO), and is furthermore unreliable
at small and possibly even intermediate momentum fractions of the emitted
parton, where the only reliable determination of FFs is via the 
Modified Leading Logarithm Approximation (MLLA) \cite{Albino:2004yg}. Despite these
problems, FFs at intermediate to large momentum fractions obtained
from fits to data now yield compatible results with other data sets \cite{Kniehl:2000fe}.

Much precise data from $e^+ e^-$ colliders now exists for the
production of the three lightest charged hadrons, which are the pion
($\pi^{\pm}$), kaon ($K^{\pm}$) and proton ($p/\overline{p}$). In much
of this data, the observed hadron is identified as one of these
particles, and the emitting parton is identified as either a gluon,
light ($u$, $d$ and $s$) quark, $c$ quark or $b$ quark, which allowed
for a precise determination of the corresponding individual FFs in
Refs.\ \cite{Kniehl:2000fe,Kretzer:2000yf} \cite{footnote1}. 
However, the individual light quark FFs
could only be extracted by making reasonable physical assumptions.

Since this analysis, the OPAL Collaboration has presented light
flavour separated measurements on light charged hadron production
\cite{Abbiendi:1999ry} for the $e^+ e^-$
centre-of-mass (CM) energy $\sqrt{s}=M_Z$, allowing for the first time the
extraction of flavour dependent FFs of light quarks. In this OPAL analysis, high energy
mesons ($\pi^{\pm}$, $K^{\pm}$ and $K_S^0$) and baryons ($p/\overline{p}$ and $\Lambda$)
were identified in the large $Z$ boson decay data sample and used as tagging products. In addition,
high momentum $e^{\pm}$, $\mu^{\pm}$ and $D^{* \pm}$ particles and identified
bottom events were used to measure heavy flavour backgrounds in the above meson and
baryon sample. As suggested in Ref.\ \cite{Field:1976ve} and precisely studied in
a recent analysis by the SLD Collaboration \cite{Abe:1998zs}, these
high energy particles give information about the original quark. For more details
see the OPAL work \cite{Abbiendi:1999ry}, where it is explained how the Collaboration 
measured the probability
$\eta_a^h(x_p,s)$ for a quark flavour $a$ to develop into a jet containing the particle $h$ with
a momentum fraction $x$ larger than $x_p=2p_h/\sqrt{s}$.

Since the valence structure of the proton is $uud$, knowing the difference between the 
individual light flavour FFs, in particular for
$u$ and $d$ quarks into $K^{\pm}$, is very much needed for predicting the
inclusive cross sections for the productions of these hadrons in
collisions involving protons, such as $ep$, $pp$ and $p\overline{p}$
collisions. For example, results from the inclusive production of hadrons in 
$pp$ collisions provide the
baseline to which one compares heavy-ion collision results in order to
determine the properties of the hot quark-gluon plasma \cite{d'Enterria:2004rp}.
Tests presented in Ref.\ \cite{Kniehl:2000hk} of the KKP FFs in the process
$p +\overline{p}\rightarrow h^{\pm}+X$, where $h^{\pm}$ are light
charged hadrons, were generally successful, as was a recent check of
the pion FFs by comparison to $p+p\rightarrow \pi^0 +X$ data
(taking $\pi^0=\frac{1}{2}\left(\pi^+ +\pi^-\right)$) from the 
PHENIX Collaboration \cite{Adler:2003pb} at RHIC.
However, it is likely that the inaccuracy on the information on the $u$, $d$ and $s$ quark FFs
canceled out due to the superimposition of the
hadrons in $h^{\pm}$. 

In this paper, we update the analysis of Ref.\ \cite{Kniehl:2000fe} by
including the data of Ref.\ \cite{Abbiendi:1999ry} in the fit to obtain for the first
time a phenomenological determination of the individual light quark
FFs for each light charged hadron species. Since we do not impose those
physical assumptions on the light quark FFs that were used in Ref.\ 
\cite{Kniehl:2000fe} in our calculation of the cross sections used for
the fit, the other FFs extracted in this fit are also more reliable.
In Section \ref{formalism}, we summarize the basic theoretical tools used in our calculations for the fit.
In Section \ref{method} we justify specific choices for our fit such as the data used and 
the FF parameterization. Our results are then presented in Section \ref{results},
and finally in Section \ref{conclusions} we present our conclusions. The details
of the longitudinal cross section calculation are given in Appendix \ref{app}.

\section{Formalism \label{formalism}}

The optimal way to determine FFs is to fit them to measurements of the processes 
$e^+ +e^- \rightarrow (\gamma,Z)\rightarrow a +\overline{a}\rightarrow h+X$,
where $a$ is the tagged quark,
$h$ is a detected hadron and $X$ is the remaining unobserved part of
the final state. In a typical experiment the hadron is only
detected if its species $h$ belongs to a specified set of hadron species $S_H$ and the species of
the tagged quark $a$ belongs to a set of flavours $S_A$. Writing the CM momentum of the
observed hadron as $x\sqrt{s}/2$, the data for
such a process are typically presented as 
\beq F^{S_H}_{S_A}(x,s)
=\frac{\sum_{a\in S_A,\ h\in S_H}\frac{d\sigma_a^h}{dx}(x,s)}
{\sum_{a\in S_A}\sigma_a(s)}.
\label{genformofdata}
\eeq
The total cross section $\sigma_a$ is given to NLO by
\beq
\sigma_a(s)=\sigma_0(s) N_c Q_a(s) \left(1+2a_s(s)\right),
\eeq
where $\sigma_0=4\pi\alpha^2/(3s)$ is the leading order (LO) cross section for the 
process $e^+ +e^- \rightarrow \gamma \rightarrow 
\mu^+ +\mu^-$, $N_c$ is the number of colours and $a_s(\mu^2)
=\alpha_s(\mu)/(2\pi)$. $Q_a(s)$ is the effective electroweak charge of quark $a$ \cite{Eidelman:2004wy}. 

From the Factorization Theorem, the higher twist component of the 
differential cross section in Eq.\ (\ref{genformofdata}) is of $O(1/\sqrt{s})$ or less
and may and will be neglected in this paper, while the leading twist component is
obtained by convoluting the corresponding high energy partonic cross
sections with the FFs $D_a^h(y,M_f^2)$, where $y$
is the fraction of the momentum of parton $a$ taken away by the produced hadron $h$ and $M_f$
is the factorization scale. This may be written concisely by taking $y=x/z$, in which case
\beq
\begin{split}
&\frac{d\sigma^h_a}{dx}(x,s)=
\int_x^1 \frac{dz}{z}\Bigg[
\frac{d\sigma^a_{a,{\rm NS}}}{dz}\left(z,s,M_f^2\right)
D_a^h\left(\frac{x}{z},M_f^2\right)\\
&+\sum_b \frac{d\sigma^b_{a,{\rm PS}}}{dz}\left(z,s,M_f^2\right)
D_b^h\left(\frac{x}{z},M_f^2\right)
+\frac{d\sigma^g_a}{dz}\left(z,s,M_f^2\right)
D_g^h\left(\frac{x}{z},M_f^2\right)
\Bigg],
\end{split}
\label{XSfromFFs}
\eeq
where, for the emission of a hadron $h$ from a quark $a$, the non-singlet
partonic cross section $d\sigma^a_{a,{\rm NS}}/dz$ contains only and all those contributions from
diagrams in which the quark line connected to the electroweak vertex and the quark line
emitting the hadron $h$ are the same, while the pure singlet
partonic cross section $d\sigma^a_{b,{\rm PS}}/dz$ contains all other contributions.
Since the $Z$ boson only splits into a quark $a$ and its antiquark $\overline{a}$, each 
partonic cross section is proportional to $Q_a$, and thus may be written
\cite{footnote2}
\beq
\begin{split}
\frac{d\sigma^a_{a,{\rm NS}}}{dz}\left(z,s,M_f^2\right)
=&\sigma_0 (s) Q_a(s)C_{\rm NS}\left(z,a_s(s),\ln \frac{M_f^2}{s}\right),\\
\frac{d\sigma^b_{a,{\rm PS}}}{dz}\left(z,s,M_f^2\right)
=&\sigma_0 (s) \frac{Q_a(s)}{n_f}
C_{\rm PS}\left(z,a_s(s),\ln \frac{M_f^2}{s}\right)\ {\rm and}\\
\frac{d\sigma^g_a}{dz}\left(z,s,M_f^2\right)=&2\sigma_0 (s) 
Q_a(s)C_g\left(z,a_s(s),\ln \frac{M_f^2}{s}\right),
\end{split}
\eeq
where the $C_i$ are the perturbatively calculable coefficient functions.
$n_f$ is the number of active quark flavours.
For the choice $M_f^2=s$, the $C_i(z,a_s(s),0)=C_i(z,a_s(s))$ for the unpolarized (i.e.\ 
summed over transverse and longitudinal components)
cross section are given to NLO by \cite{Altarelli:1979kv}
\begin{eqnarray}
C_{\rm NS}(z,a_s)&=& \delta(1-z)+a_s C_F
\Bigg[\left(\frac{2\pi^2}{3}-\frac{9}{2}\right)\delta(1-z)
-\frac{3}{2}\left[\frac{1}{1-z}\right]_+ \nonumber \\
&& +(1+z^2)\left[\frac{\ln (1-z)}{1-z}\right]_+
+1+2\frac{1+z^2}{1-z}\ln z +\frac{3}{2}(1-z)\Bigg],\\
C_{\rm PS}(z,a_s)&=&O(a_s^2)\ \ \ {\rm and}\\
C_g(z,a_s)&=&a_s  C_F \left[
\frac{1+(1-z)^2}{z}\left(\ln (1-z)+2\ln z\right)-2\frac{1-z}{z}\right].
\end{eqnarray}
Note that the pure singlet contribution only enters at NNLO.
In contrast, in the longitudinal cross section
\beq F^{S_H}_{L,S_A}(x,s)
=\frac{\sum_{a\in S_A,\ h\in S_H}\frac{d\sigma_{L,a}^h}{dx}(x,s)}
{\sum_{a\in S_A}\sigma_a(s)},
\label{genformofLdata}
\eeq
there is a contribution from the pure singlet sector at NLO, while the 
gluon FF enters at LO (see Appendix \ref{app}). 

It is clear that we only apply electroweak theory to LO. We can therefore easily see that
Eq.\ (\ref{XSfromFFs}) for the cross section when quark $a$ is tagged
is a physical observable, since it can be obtained
by differentiating the untagged cross section (Eq.\ (\ref{XSfromFFs}) with $a$ summed over all
flavours) with respect to $\ln Q_a$, where $Q_a$
is the effective electroweak charge of quark $a$ discussed above. 
Therefore the tagged cross section is formally 
independent of the factorization and renormalization scales and
schemes, as it must be to qualify as an observable.

For $M_f^2\neq s$, the coefficient functions will contain terms of the form
$a_s^n(s) \ln^p \frac{M_f^2}{s}$, where $p=n,n-1,...$, which will spoil the convergence
of the series unless $M_f^2=O(s)$. Thus, in order to be able to describe data over a large
range in $s$, the dependence of the FFs on $M_f^2$ must be known. Fortunately this can
be calculated using the DGLAP equation,
\beq
\frac{d}{d \ln M_f^2} D_a^h(z,M_f^2)=\sum_{b=g,q} 
\int_z^1 \frac{dy}{y}P_{ab}\left(\frac{z}{y},a_s(M_f^2)\right)
D_b^h(y,M_f^2),
\label{DGLAPeq}
\eeq
where the $a\rightarrow b$ splitting functions $P_{ab}$ 
are perturbatively calculable, and are known to NLO.
Therefore, in the calculation of cross sections
it is sufficient to know the FFs at just one factorization
scale $M_f=M_0$.

The DGLAP equation is however not valid when $z$ is small, since due to soft gluon emission
the $P_{ag}(z,a_s)$ contain terms which behave in the limit $z\rightarrow 0$ like
$(a_s^n/z) \ln^{2n-1-m}z$, where
$m=1,...,2n-1$ labels the class of terms (finite terms which behave like $a_s^n$ when 
integrated over the range $0<z<1$
are classified as $m=2n$), and are therefore unreliable in this limit. This implies that
the cross section cannot be reliably calculated at small $x$, and the FFs $D_a^h(z,M_0^2)$
cannot be fitted at small $z$. In this case a description
of the data requires an alternative approximation
such as the MLLA, which is beyond the scope of this paper.

Dependence on the factorization scale is introduced
in the usual way. Specifically, the FFs are 
evolved to $M_f^2=k_f s$, where $k_f$ is a constant which is taken
to be equal to 1 for the main fit, and 1/4 and 4 in two further fits to determine
the theoretical errors on fitted parameters. We counter-balance this $k_f$ dependence at NLO 
using the result (where the 
$x$ dependence, integrals, discrete labels, sums and charges have been removed for brevity)
\beq
C(a_s(s),\ln k_f)=C(a_s(s),0)-\ln k_f C (a_s(s),0) P (a_s(s)).
\label{coeffwithkfnotone}
\eeq
Dependence on the renormalization scale $\mu$ is introduced by choosing $\mu^2=k s$, where $k$
is a constant chosen to obey $k=k_f$. At NLO, this amounts to replacing $a_s(s)$ in the
coefficient functions with $a_s(k s)$.

The fastest and most accurate way of calculating a cross section is 
in Mellin space, defined by the transformation 
\beq
F^{S_H}_{S_A}(n,s)=\int_0^1 dx x^{n-1} F^{S_H}_{S_A}(x,s),
\eeq
since convolutions such as that in Eq.\ (\ref{XSfromFFs}) become simple products.
In particular, Eq.\ (\ref{DGLAPeq}) becomes
\beq
\frac{d}{d \ln M_f^2}D_a^h(n,M_f^2)=\sum_{b=g,q} P_{ab}(n,a_s(M_f^2))
D_b^h(n,M_f^2),
\eeq
which can be solved analytically
order by order. The cross section in $x$ space can then be obtained numerically
via the inverse Mellin transform,
\beq
F^{S_H}_{S_A}(x,s)=\frac{1}{2\pi i}\int_C dn\ x^{-n} F^{S_H}_{S_A}(n,s),
\label{invmeltran}
\eeq
where $C$ is a contour in Mellin space from ${\rm Im} (n)=-\infty$ to ${\rm Im} (n)=\infty$, which
passes to the right of all poles.

Predictions for data averaged over an $x$-bin in the range $x_l<x<x_h$ are calculated from the formula
\beq
\langle F^{S_H}_{S_A}\rangle (x_l,x_h,s)=\frac{1}{x_h-x_l}\int_{x_l}^{x_h}dx 
F^{S_H}_{S_A}(x,s).
\eeq
This integral over $x$ can be done analytically in
Eq.\ (\ref{invmeltran}), 
\beq
\langle F^{S_H}_{S_A}\rangle (x_l,x_h,s)=\frac{1}{x_h-x_l}
\frac{1}{2\pi i}\int_C dn \frac{x_h^{1-n} -x_l^{1-n}}{1-n} F^{S_H}_{S_A}(n,s),
\label{xintinvmelltrans}
\eeq
giving a further advantage for
working in Mellin space that no extra numerical integration is required to obtain
$x$-bin averaged cross sections.

The light flavour separated data in Ref.\ \cite{Abbiendi:1999ry} may be interpreted as the probability
for a tagged quark flavour $a$ to inclusively emit a hadron of type $h$ with momentum greater than
$x_p \sqrt{s}/2$, in which case the corresponding
theoretical result for such data may be calculated from the formula
\beq
\eta^h_a(x_p,s)=\int_{x_p}^{1}dx F^{\{h\}}_{\{a\}}(x,s)=
(1-x_p)\langle F^{\{h\}}_{\{a\}}\rangle (x_p,1,s),
\label{calcofetas}
\eeq
and we note that for this expression the $\eta^h_a(x_p,s)$ constrain the 
FFs at large momentum fraction even more than the $F^{S_H}_{S_A}(x,s)$.
However, the experimental definition of the $\eta^h_a$ is a little more subtle.
For a given number $N_a$ of $e^+e^-$ annihilation events in which a quark $a$ is tagged, 
the number $N_{a\rightarrow h}$ of times that
an {\it event hemisphere}, defined to be the two regions separated
by the plane perpendicular to the thrust axis for each event,
contains a particle $h$ with $x>x_p$ is determined. Therefore,
at LO, where $a$ and $\overline{a}$ are never in the same hemisphere,
$\eta_a^h (x_p,s)$ is given by the integral over $D_a^h(x,s)$ in the range $x_p<x<1$,
and this result is consistent with Eq.\ (\ref{calcofetas}). At NLO
the quark $a$ can emit a gluon which in turn emits the hadron $h$ according to the
gluon FF $D_g^h$ (see Eq.\ (\ref{XSfromFFs})). 
In the measurement of $\eta^h_a(x_p,s)$, processes in which the gluon is 
in the opposite hemisphere from the quark $a$ that emitted it are excluded.
However, such processes contribute to Eq.\ (\ref{calcofetas}). Fortunately,
such events in which the gluon is emitted with a large angle with respect to the quark $a$ are very
rare and should contribute very little both
to Eq.\ (\ref{calcofetas}) and the measured $\eta^h_a$.

\section{Method \label{method}}

In this Section we describe our method for obtaining FFs from data. 
As in Ref.\ \cite{Kniehl:2000fe}, where a detailed discussion of all available 
data sets is given which 
will not be repeated here, we use identified hadron data with and without flavour
separation from DELPHI \cite{Abreu:1998vq} and SLD \cite{Abe:1998zs}, 
and identified hadron data without flavour separation from 
ALEPH \cite{Buskulic:1994ft} and TPC \cite{Aihara:1988fc}. 
In addition, we use identified hadron data with flavour separation
from TPC \cite{Aihara:1986mv}, which was used in Ref.\ 
\cite{Kretzer:2000yf} but not in Ref.\ \cite{Kniehl:2000fe}.
Furthermore, for the first time we also include the light
flavour separated measurements of quark tagging probabilities from the
OPAL Collaboration \cite{Abbiendi:1999ry}. 
However, we exclude unidentified hadron data since, although such data is accurate, it is
typically contaminated with charged particles other than the
$\pi^{\pm}$, $K^{\pm}$ and $p/\overline{p}$. Such data was used in
Ref.\ \cite{Kniehl:2000fe}, leading to consistent results. However, since in
this analysis we aim for more reliable FFs, we use only hadron species
separated measurements. We also exclude data for which $x_l<0.1$, since
the prediction for the cross section is unreliable in this region as a result of 
the logarithms from soft gluon emission mentioned in Section \ref{formalism}.
After fitting, we then compare
cross sections calculated from our FFs and $\alpha_s(M_Z)$ with
the unidentified hadron data with flavour separation
from TPC \cite{Aihara:1986mv}, with and without flavour separation
from ALEPH \cite{Buskulic:1995aw,PadillaAranda:1995wi}, DELPHI \cite{Abreu:1998vq} and
OPAL \cite{Ackerstaff:1998hz}, without flavour separation from
SLD \cite{Abe:1998zs}, the unidentified
hadron gluon-tagged three-jet data from ALEPH \cite{Barate:1998cp} and OPAL \cite{Abbiendi:1999pi}
and the identified hadron tagging probabilities 
with heavy quark flavour separation from OPAL \cite{Abbiendi:1999ry}. The latter
data set is not included in the fit since the heavy quark FFs are much better
constrained by the larger quantity and quality of heavy quark-tagged data from DELPHI, SLD and TPC.
In all data, correlation effects between data points are not yet known, and therefore
in the calculation of 
the covariance matrix for the $\chi^2$ to be minimized, we fix the off-diagonal
elements to zero, i.e.\ we assume that the data points are uncorrelated and that the
error on each one is given by its statistical and systematic errors added in quadrature.
Note that this common deficiency in published data limits the
reliability of results obtained from analyzing them.

All theoretical quantities are calculated to NLO in the $\overline{\rm MS}$ scheme.
For our main fit, we evolve the FFs from $M_f=M_0$ to 
$M_f=\sqrt{s}$, and vary the scales as described in Section
\ref{formalism} to determine the theoretical errors on $\alpha_s(M_Z)$.
We take $M_0=\sqrt{2}$ GeV for $a=u,d,s,g$. As $M_f$ is increased
from $M_0$ to $\sqrt{s}$, the number of
flavours used in the evolution of the FFs and the strong coupling is first set to $n_f=3$ and only
the light quark and gluon FFs are non zero until $M_f=m(\eta_c)=2.9788$ GeV, where the charm FF is
set equal to its initial distribution and included in the set of FFs to be evolved, and the number
of flavours is taken to be $n_f=4$. The bottom FF is treated in the same way,
being introduced when $M_f=m(\Upsilon)=9.46037$ GeV. 
Both flavour thresholds are respectively twice the pole masses of these two heavy quarks,
and therefore perturbative matching conditions are required at NLO. Rather than
implementing this matching explicitly, we define our heavy quark FFs to be the complete ones,
not just the intrinsic FFs, which means the matching term, dependent
on the gluon FF, is absorbed into them.
Our FFs are summed over hadrons which are of the same species but
opposite charges, and averaged over quark and antiquark. We do not 
consider cross sections which depend on the difference between
quark and antiquark FFs summed over any given set of emitted hadrons, although it must
be noted that this difference is zero when this set contains a sum over charges, by
charge conjugation invariance. 
Since we use accurate data at 29 GeV and $91.2$ GeV, we are in a
position to extract the parameter $\alpha_s(M_Z)$, the quantity
which determines the running of $a_s(\mu^2)$. We therefore
free this parameter in our fit. 
The matching on $a_s(\mu^2)$ is implemented by determining 
$\Lambda_{\overline{\rm MS}}^{(5)}$ from $\alpha_s(M_Z)$, and then using it to determine
$\Lambda_{\overline{\rm MS}}^{(4)}$ and $\Lambda_{\overline{\rm MS}}^{(3)}$ from the NLO relations given in
Ref.\ \cite{Marciano:1983pj} (these were checked using the results of
Ref.\ \cite{Chetyrkin:1997sg}). 
We choose the usual parameterization
\beq
D_a^h(x,M_0^2)=N x^{\alpha} (1-x)^{\beta}
\label{param}
\eeq
for each of our FFs. In Mellin space, the FFs are then proportional to
$\Gamma(n+\alpha)/\Gamma(n+\alpha+\beta+1)\simeq 1/(n+\alpha)$ for $n\simeq -\alpha$, 
and this behaviour persists
even after evolution and convolution with coefficient functions. 
For such behaviour, the numerical evaluation of Eq.\ (\ref{xintinvmelltrans}) is
best performed with the integration variable $0\leq t\leq 1$ and contour defined through
\beq
n=c+\frac{3}{3+\ln\frac{1}{x_l}}+\frac{1}{1-x_l}+\frac{2(1-i)\ln t}{\ln\frac{1}{x_l}},
\label{nfromt}
\eeq
where the real constant $c$ is chosen such that the contour lies to the right
of all poles, 
since as $t \rightarrow 1$ the integrand in the integral over $t$ becomes a finite
constant, while as $t \rightarrow 0$ the integrand vanishes like 
$\exp ((-2(1-i)\ln t / \ln (1/x_l))\ln x)$. As a result of the 
second and third term in Eq.\ (\ref{nfromt}),
the intersection of the contour with the ${\rm Im}(n=0)$ line goes from $n=1$
to $n=\infty$ as $x$ goes from 0 to 1. This approximately follows the saddle point \cite{RDBall}
of the integrand,
thus ensuring the contour is close to the contour of steepest descent, which gives the fastest
convergence of the integral.

In Ref.\ \cite{Kniehl:2000fe}, no data was used which could allow
for the difference between the $d$ and $s$ FFs to be determined. 
(The FFs for the $u$ can be determined since its electroweak charge is
different to that of $d$ and $s$.) The authors 
constrained this difference by imposing
the valence quark structure at all momentum fractions and SU$(3)$
invariance, giving the relations
\beq
\begin{split}
D_u^{\pi^{\pm}}(x,M_0^2)&=D_d^{\pi^{\pm}}(x,M_0^2),\\
D_u^{K^{\pm}}(x,M_0^2)&=D_s^{K^{\pm}}(x,M_0^2)\ {\rm and}\\
D_u^{p/\overline{p}}(x,M_0^2)&=2D_d^{p/\overline{p}}(x,M_0^2).
\end{split}
\label{lqconstraints}
\eeq
Such constraints can be implemented by fixing the parameters $N$, $\alpha$ and $\beta$ of the FFs
on the right hand sides to be equal to those of the FFs on the left hand
side, with the exception that the parameter $N$ of $D_u^{p/\overline{p}}$ must be fixed to twice 
the value of that of $D_d^{p/\overline{p}}$
\cite{footnote3}.
With such conditions on the parameterization, a good fit to the data used was obtained. 

The first line in Eq.\ (\ref{lqconstraints}) also follows from SU$(2)$
isospin invariance, and is therefore expected to be accurate
\cite{Gronau:1973gc}. Indeed, the approximate result $\eta_d^{\pi^{\pm}}=\eta_u^{\pi^{\pm}}$
implied by this relation is found to hold within
$2\%$ for $x_p\geq 0.2$. However, the second line in Eq.\ 
(\ref{lqconstraints}) is expected to be strongly violated since the
$s$ quark has a significantly larger mass than the $u$ quark. Already
in 1977, Field and Feynman \cite{Field:1976ve} assumed that due to the
larger mass of $s$ quarks, the $\overline{s} \rightarrow K^+$ transition
should happen more frequently than the $u \rightarrow K^+$ one because
less energy is needed for the creation of a $u\overline{u}$ pair from the
vacuum than for a $s\overline{s}$ pair. This is measured by the suppression
factor $\gamma_s$ of strange quarks, which is known from various
strange/non-strange hadron production rates to be around
$\gamma_s\simeq 0.3$. (For a compilation, see Ref.\ \cite{Knowles:1995kj}.)
The third line in Eq.\ (\ref{lqconstraints}), assumed earlier also in
Ref.\ \cite{Baier:1979tp}, can also be justified for $x\rightarrow 1$ by the valence
ratios and dimensional counting powers \cite{Jones:1978he}. Indeed, in the OPAL
analysis of Ref.\ \cite{Abbiendi:1999ry}, the ratio $\eta_d^{p/\overline{p}}/\eta_u^{p/\overline{p}}$
is consistent with 0.5 for all $x_p \geq 0.2$, but only inside the rather large errors.
However, decays from heavier baryons such as $\Lambda$ or $\Delta$ resonances might change this ratio. 
Furthermore, within the LUND string model \cite{Andersson:1983ia} the actual value of the ratio 
$\eta_d^{p/\overline{p}}/\eta_u^{p/\overline{p}}$
at large $x_p$ would be a direct measure of the size of the
suppression of diquarks with spin 1 relative to those with spin 0,
since Fermi-Dirac statistics requires a $uu$ diquark to have angular
momentum $L=1$. In summary, all relations in Eq.\ (\ref{lqconstraints}), particularly
the last two, may be violated to a possibly relevant degree, but in any case
since we will use the data of Ref.\ \cite{Abbiendi:1999ry} in our analysis,
we shall not impose any relations between the light quark flavour FFs.

\section{Results \label{results}}

In this Section we report the results obtained from the fit described in
Section \ref{method}. We obtain
\beq
\alpha_s(M_Z)=0.1176^{+0.0053}_{-0.0067}
{\rm [exp]}^{+0.0007}_{-0.0009}{\rm [theo]}=
0.1176^{+0.0053}_{-0.0068}.
\label{resforalphas}
\eeq
This is equivalent to the result
$\Lambda_{\overline{\rm MS}}^{(5)}=221\pm 74{\rm [exp]}_{-10}^{+9}{\rm [theo]}$ MeV.
The experimental errors are obtained by varying $\alpha_s(M_Z)$, keeping all other parameters fixed,
until $\chi^2_{\rm DF}$ increases by unity. 
The theoretical errors, determined using the method described in Section
\ref{formalism}, turn out to be negligible relative to the experimental ones, 
most likely because the $x$ range of the data used is very limited.
The second result in Eq.\ (\ref{resforalphas}), whose upper and lower errors are obtained by
adding the upper and lower errors respectively of both sources in quadrature, 
is consistent with the KKP result \cite{Kniehl:2000cr} of $\alpha_s(M_Z)=0.1170^{+0.0058}_{-0.0073}$
(which includes the theoretical error).
In Table \ref{pars}, we show the values of the remaining, FF parameters obtained from the fit.
Since $N$ and $\beta$ are highly correlated and the large $x$ data generally
has the largest errors, for some FFs these two parameters are large. However, 
over the range $0.1\leq x \leq 1$, all FFs are of similar order in magnitude.
Also shown in Table \ref{pars} is the symmetrized propagated experimental error on each parameter. 
This quantity is the average of the two resulting errors 
obtained by varying the parameter, keeping the other parameters fixed, 
until $\chi^2$ increases by 1 from its minimum value. The
correlated errors
between the parameters are expected to be of a
similar order of magnitude to the purely statistical errors shown.
Note that these results show no obvious consistency with Eq.\ (\ref{lqconstraints}).
With the inclusion of correlation effects in the data, a deeper investigation into parameter
errors would be worthy.

We obtained $\chi^2_{\rm DF}=1.15$, indicating an
overall good description of the data. 
The resulting $\chi^2_{\rm DF}$ values for the OPAL light quark
tagging probabilities from Ref.\ \cite{Abbiendi:1999ry}
are shown in Table \ref{chiDFforlqetas}. The description of the data in which $K^{\pm}$ or
$p/\overline{p}$ is detected is excellent, except for the process
$d\rightarrow p/\overline{p}$. For this and the $\pi^{\pm}$ data, which has the highest accuracy,
a fit without the data points at $x_p=0.2$
results in all values of $\chi^2_{\rm DF}$ being around
unity, although the resulting FFs from that fit are not considerably different to those from
the main fit. This data, together with
the corresponding theoretical curves calculated from our FF set (labeled AKK), and with the curves
from the sets of Ref.\ \cite{Kniehl:2000fe} (labeled KKP) and 
Ref.\ \cite{Kretzer:2000yf} (labeled Kretzer), are shown in Fig.\ \ref{fig1}. We see 
that for the $s,d\rightarrow K^{\pm}$ transitions, the corresponding AKK curves are in good
agreement with the data while the KPP and Kretzer curves strongly disagree. The Kretzer set
fails to lead to a decent description
for the $\eta^{\pi}_d$ data, but otherwise all $\pi^{\pm}$ data is well described
by all three sets. Our set and the KKP set lead to a good description of the $p/\overline{p}$
data (which were not used in the determination of the Kretzer set).
Fig.\ \ref{fig7} shows the heavy quark tagging probabilities, which were not used in the fit,
together with the corresponding theoretical curves from the same FF sets as were used in 
Fig.\ \ref{fig1}. 
In Table \ref{chiDFforhqetas} we list the corresponding $\chi^2_{\mathrm{DF}}$ values. Clearly
these values are unacceptably high.
In order to check that this was not a result
of the inadequacy of our parameterization to allow for a description of both 
small $x$ and large $x$ data (since, as discussed around Eq.\ (\ref{calcofetas}), 
the OPAL quark tagging probabilities provide
more constraints on the FFs at large $x$), we performed three new fits which
included the heavy quark tagging probabilities, the first being otherwise similar to the main fit,
the other two having the following differences:
For the second fit, the quark FFs were modified by 
multiplying the right hand side of Eq.\ (\ref{param}) with $(1+\gamma x)$, with
$\gamma$ different for each quark FF and fixed to zero for the gluon FF, and each $\gamma$ 
was included in the set of free parameters to be fitted.
In the third fit, all $x_l<0.2$ data were excluded. No significant improvement to
the description of the heavy quark tagging probabilities was obtained in all three fits.
We therefore assume that this discrepancy is caused by the inclusion of
large angle gluon emission effects in Eq.\ (\ref{calcofetas}), as described at the end of Section 
\ref{formalism}. However, since we have sufficient data to constrain
the heavy quark FFs, we will not pursue this problem further in this paper.
All remaining values of $\chi^2_{\rm DF}$ from data used in the fit are listed in Table \ref{allchis}.
Each of these lie around or below unity.
Since an excellent fit is obtained to DELPHI, SLD and TPC heavy quark-tagged data,
we conclude that our fitted heavy quark FFs are reliable even though 
using them in Eq.\ (\ref{calcofetas}) leads to a poor description of 
the OPAL heavy quark tagging probabilities. Since the DELPHI, SLD and TPC
light quark-tagged data is well fitted with the light quark 
tagging probabilities, Eq.\ (\ref{calcofetas}) is sufficient
for describing the latter data.
The values of $\chi^2_{\rm DF}$ for the data to be used for comparison, which were 
discussed at the beginning of Section \ref{method}, are also shown. 
The serious disagreement with the ALEPH 
\cite{Buskulic:1995aw,PadillaAranda:1995wi} 
and OPAL \cite{Ackerstaff:1998hz} data found here was also found in Ref.\ \cite{Kniehl:2000fe},
where it was argued that this data has a sizeable contribution from
charged particles other than the three lightest charged hadrons. For the ALEPH
data without flavour separation, this argument is supported by the fact 
that the data for charged hadron production significantly overshoots the sum
of the hadron identified data.

In Figs.\ \ref{fig2} -- \ref{fig5}, we show all these normalized differential cross section
data used for fitting and for comparison,
together with the corresponding theoretical curves from the fit. 
The TPC flavour separated data \cite{Aihara:1986mv}, particularly the $uds$ quark-tagged data, lie 
far from their theoretical predictions. However, it must be understood that these data are rather old
compared to the rest of the data used in the fit. At any rate, using them
has not affected the overall quality of the fit since their errors are large, which
explains why their $\chi^2_{\rm DF}$ values in Table \ref{allchis} are not too far from unity.
Qualitatively, at least, the rise in the calculated cross section at low $x$ for decreasing $\sqrt{s}$
is confirmed by the TPC data, as was first noted in Ref.\ \cite{Kretzer:2000yf}.
These figures show that the only TPC data which 
can significantly constrain $\alpha_s(M_Z)$ are the $\pi^{\pm}$ and $K^{\pm}$ identified data
shown in Fig.\ \ref{fig2}. In Fig.\ \ref{fig6}, we show the gluon-tagged three-jet data together with the 
theoretical curves for $D_g(x,4 E_{\rm jet}^2)$. The resulting $\chi^2_{\mathrm{DF}}$ values
shown in the last two lines of Table \ref{allchis} are 
very high, but it must be kept in mind that the theoretical calculation is only correct at LO, and
the gluon is only determined at LO.
In Ref.\ \cite{Kniehl:2000fe}, where this data was used in the fit, 
this identification was made only because the gluon FF is 
much less constrained by the remaining data than the quark FFs.

In Fig.\ \ref{fig8}, we compare the longitudinal cross section with the data
without flavour separation from 
ALEPH \cite{Buskulic:1995aw}, DELPHI \cite{Abreu:1997ir} and OPAL \cite{Akers:1995wt} 
and for light and $b$ quark separation from DELPHI \cite{Abreu:1997ir}.
The $x$ space coefficient functions of the longitudinal cross section
are given in Ref.\ \cite{Rijken:1996vr}. However, since our cross sections
are calculated in Mellin space, we calculate the Mellin transform of these
quantities as detailed in Appendix \ref{app}. (An alternative procedure would be to evolve the FFs
in Mellin space as before, and perform the convolution of the coefficient functions
with the evolved FFs in $x$ space. However, this procedure is numerically very slow.)
In the unpolarized cross sections used in our fit, the gluon FF for each hadron enters only at NLO
and so is only determined to LO in our analysis, while it
enters at LO in the longitudinal cross section, for which a gluon FF determined to NLO
is therefore required. Thus the curves in Fig.\ \ref{fig8}
are not completely NLO, but serve to determine the quality of our gluon FF. The agreement
is excellent for the ALEPH and OPAL data, and good for the DELPHI data. Our curves are
also very similar to those obtained in Ref.\ \cite{Kniehl:2000fe}, where the LO curves 
from these authors' LO analysis are also shown. These latter curves do not agree with
the ALEPH and OPAL data as well as the NLO ones. Thus treating the LO gluon FF obtained
from their and our fits as NLO results in no loss of consistency in this case.

Finally, we compare cross sections calculated using our FFs 
for particle production in proton-(anti)proton initiated processes with experimental data. 
Such processes are highly dependent on the individual light quark flavour FFs, due
to the partonic structure of the proton.
We use the coefficient functions for the processes 
$a+ b \rightarrow c + X$, where $a$, $b$ and $c$ denote partons, 
to NLO as calculated in Ref.\ \cite{ACGG}. We convolute these with our evolved
FFs for parton $c$, and the evolved CTEQ6M parton distribution functions \cite{Stump:2003yu} 
for $a$ and $b$.
Since our fitted result of $\Lambda_{\overline{\rm MS}}^{(5)}=221$ MeV is very similar to the 
result of $\Lambda_{\overline{\rm MS}}^{(5)}=226$ MeV obtained in Ref.\ \cite{Stump:2003yu}, we use
the former result in the calculation of $a_s(\mu^2)$. We take $M_f^2=k p_T^2$. The cross section
at $x_T=2p_T/\sqrt{s}$ depends on the FFs for the whole region $x_T <z<1$.
Since we do not (reliably) determine the FFs below $z=0.1$ and/or $M_f=M_0$, we take them 
in this region to be equal to their values at this point. Graphically, we found no discernible
difference between the resulting predictions and those obtained when
the FFs in this region were fixed to zero.
Firstly, we calculate the invariant differential cross section for inclusive $\pi^0$ production for
the process $p+ p \rightarrow \pi^0 +X$ as measured by PHENIX at $\sqrt{s}=200$ GeV 
in Ref.\ \cite{Adler:2003pb}. For this we assume the relation
\beq
D_a^{\pi^0}(x,M_f^2)=\frac{1}{2}D_a^{\pi^{\pm}}(x,M_f^2)
\eeq
to be true, which follows from SU(2) flavour symmetry for pions 
(see Ref.\ \cite{Binnewies:1995pt}). Here, $D_a^{\pi^0}$ is the average of the FFs
for the processes $a,\overline{a}\rightarrow \pi^0$. (Recall $D_a^{\pi^{\pm}}(x,M_f^2)$
is also averaged over $a$ and $\overline{a}$, but summed over $\pi^+$ and $\pi^-$.)
The results are shown in Fig.\ \ref{fig9} for $k=$1/4, 1 and 4, 
together with the PHENIX data. In addition, we also compare the cross section calculated from
the FFs obtained in Ref.\ \cite{Kniehl:2000fe}. For $p_T>7$ GeV,
the curve for $k=1$ lies closer to the centre of the data than the KKP curve does.
Secondly, we calculate the invariant differential cross section for inclusive $K^0_S$ production for
the process $p+p \rightarrow K^0_S +X$ as preliminarily measured by STAR at
$\sqrt{s}=200$ GeV \cite{MH}\cite{footnote4}, and for
the process $p+\overline{p} \rightarrow K^0_S +X$ as measured by UA1 at
$\sqrt{s}=630$ GeV in Ref.\ \cite{Bocquet:1995jq}. For this we assume the relation
\beq
D_a^{K^0_S}(x,M_f^2)=\frac{1}{2}D_b^{K^{\pm}}(x,M_f^2)
\label{K0SfromKcharged}
\eeq
to be true, where $b=u,d$ if $a=d,u$, otherwise $b=a$. 
Eq.\ (\ref{K0SfromKcharged}) follows from SU(2) flavour symmetry for kaons 
(see Ref.\ \cite{Binnewies:1995pt}), and is confirmed by the fact that 
the OPAL measurements in Ref.\ \cite{Abbiendi:1999ry} for the production of
$K^0_S$ and $K^{\pm}$ mesons agree within their errors.
The predictions are shown in Fig.\ \ref{fig10}, in a format
similar to Fig.\ \ref{fig9}. For $p_T>1.5$ GeV, the $k=1$ curve agrees better
with the STAR data than the KKP curve. This disagreement in the latter case was observed in
Ref.\ \cite{MH}. However, for the older UA1 data our predictions differ considerably over the whole range,
although they are consistent with the data within the theoretical errors for $p_T>4.5$ GeV, while
the KKP curve gives good agreement. 
Since the most important difference between our analysis and
that of Ref.\ \cite{Kniehl:2000fe} is the inclusion of the OPAL tagging
data in our fit, we conclude that this agreement of the KKP curve is accidental.

\section{Conclusions \label{conclusions}}

This work is an update of the KKP analysis \cite{Kniehl:2000fe}, the
main difference being that the OPAL results on light quark tagging
probabilities have been used to phenomenologically constrain the
individual light quark FFs for the first time.  We find that the
inclusion of this data in the fit makes an important difference to the
description of the $d,s\rightarrow K^{\pm}$ transitions. 
Light flavour separated FFs are essential for making predictions
for inclusive cross sections in which there is at least one proton in the initial state
and one light hadron in the final state (or more than one, in which case 
other non perturbative quantities are also required for subprocesses in which 
multiple hadrons are emitted from
a single parton). Such cross sections will be measured, for example, at the LHC.
In addition, we have included the flavour separated TPC data \cite{Aihara:1986mv} at $\sqrt{s}=29$ GeV,
but such data makes little difference to the fit.
We have excluded all charged data to be confident that none of the data sets used were contaminated
with charged particles other than the three lightest charged hadrons. 
However, good agreement with much of the available charged hadron data, in particular
that from DELPHI and SLD, was achieved.
We point out that although our gluon FF for each hadron has been formally determined to LO only,
treating it as NLO leads to good agreement with the measured longitudinal cross sections in 
the literature. Finally, relative to the KKP predictions, we obtain
with our FFs a shift towards 
the PHENIX data for the 
invariant differential cross section for inclusive $\pi^0$ production and 
towards the STAR data for the 
invariant differential cross section for inclusive $K^0_S$ production. 

A determination of $\alpha_s(M_Z)$ has been performed. We have also calculated the theoretical error 
and find it to be negligible relative to the experimental error. 
We obtain $\alpha_s(M_Z)=0.1176^{+0.0053}_{-0.0068}$, which agrees with the
Particle Data Group's world average of $\alpha_s(M_Z)=0.1187\pm 0.002$ \cite{Eidelman:2004wy}.

In order to make predictions, our fitted FFs over the range $0.1<z<1$ and $M_0<M_f<200$ GeV 
can be obtained from the FORTRAN routines at \verb!http://www.desy.de/~simon/AKK2005FF.html!, 
which are calculated using cubic spline
interpolation on a linear grid in $(z,\ln M_f^2)$. 

\appendix
\section{APPENDIX \label{app}}

In this appendix, we give all information needed to
calculate the longitudinal coefficient functions to NLO in Mellin
space.

The coefficient functions for the longitudinal cross section are given to NLO by \cite{Rijken:1996vr}
{\footnotesize
\beq
\begin{split}
&C_{\rm L,NS}(z,a_s)=a_s C_F+a_s^2 \Bigg[C_F^2\bigg{\{}4 S_{1,2}(1-z)-12{\rm Li}_3(-z)
+4\ln z {\rm Li}_2(-z)\\
&+4\zeta(2)\ln(1+z)+4\left[2\ln(1+z){\rm Li}_2(-z)+\ln z \ln^2(1+z)+2S_{1,2}(-z)\right]\\
&-4\zeta(2)\ln(1-z)-2\ln^2 z \ln(1+z)-4\zeta(3)+\left(\frac{7}{2}+z\right)\ln (1-z)\\
&+\left(-4+\frac{12}{5}z^{-2}-8z-\frac{8}{5}z^3\right)
\left[-\frac{1}{2}\widetilde{\Phi}(z)+\frac{1}{4}\ln^2 z-\frac{\zeta(2)}{2}\right]
-3{\rm Li}_2(1-z)\\
&+\ln z \ln(1-z)+\left(4-8z-\frac{8}{5}z^3\right)\zeta(2)+\ln^2(1-z)
+\left(-\frac{3}{2}+4z+\frac{4}{5}z^3\right)\ln^2 z\\
&+\left(\frac{17}{10}
-\frac{12}{5}z^{-1}+\frac{6}{5}z+\frac{8}{5}z^2\right)\ln z
-\frac{147}{20}+\frac{12}{5}z^{-1}-\frac{9}{10}z+\frac{8}{5}z^2\bigg{\}}\\
&+C_A C_F\bigg{\{}-2S_{1,2}(1-z)
-2\left[2\ln (1+z){\rm Li}_2(-z)+\ln z \ln^2(1+z)+2S_{1,2}(-z)\right]\\
&+6{\rm Li}_3(-z)-2\zeta(2)\ln (1+z)+2\zeta(2)\ln(1-z)-2\ln z {\rm Li}_2(-z)+\ln^2 z\ln (1+z)\\
&+\left(2-\frac{6}{5}z^{-2}+4z+\frac{4}{5}z^3\right)
\left[-\frac{1}{2}\widetilde{\Phi}(z)+\frac{1}{4}\ln^2 z-\frac{\zeta(2)}{2}\right]
+\zeta(2)\left(4z+\frac{4}{5}z^3\right)\\
&-\left(2z+\frac{2}{5}z^3\right)\ln^2 z
-\frac{23}{6}\ln(1-z)+\left(-\frac{73}{30}+\frac{6}{5}z^{-1}+\frac{2}{5}z-\frac{4}{5}z^2\right)\ln z
+2\zeta(3)\\
&+\frac{1729}{180}-\frac{6}{5}z^{-1}-\frac{49}{30}z-\frac{4}{5}z^2\bigg{\}}\\
&+C_F T_R n_f\bigg{\{}\frac{2}{3}\left(\ln (1-z)+\ln z\right)-\frac{25}{9}+\frac{2}{3}z\bigg{\}}\Bigg],
\end{split}
\eeq
\beq
\begin{split}
&C_{\rm L,PS}(z,a_s)=a_s^2 C_F T_R n_f\bigg{\{}4{\rm Li}_2(1-z)+4\ln z \ln (1-z)+6\ln^2 z
-\frac{28}{3}-4z^{-1}\\
&+\frac{52}{3}z-4z^2+\left(\frac{8}{3}z^{-1}-4z+\frac{4}{3}z^2\right)\ln(1-z)
+\left(-8+\frac{16}{3}z^{-1}-8z+\frac{4}{3}z^2\right)\ln z
\bigg{\}}
\end{split}
\eeq
and
\beq
\begin{split}
&C_{\rm L,g}(z,a_s)=a_s C_F\left[\frac{2}{z}-2\right]
+a_s^2 \Bigg[C_F^2 \bigg{\{} 
+2{\rm Li}_2 (1-z)+2 \ln z \ln (1-z)+\frac{4}{15}\zeta(2)z^3\\
&+\left(3-\frac{2}{15}z^3\right)\ln^2 z+\left(-\frac{4}{3}+\frac{8}{5}z^{-2}+\frac{4}{15}z^3\right)
\left[-\frac{1}{2}\widetilde{\Phi}(z)+\frac{1}{4}\ln^2 z-\frac{\zeta(2)}{2}\right]\\
&+\left(-3+4z^{-1}-2z\right)\ln (1-z)+\left(-\frac{1}{5}+\frac{12}{5}z^{-1}-\frac{28}{15}z
-\frac{4}{15}z^2\right)\ln z \\
&+\frac{3}{5}-\frac{12}{5}z^{-1}+\frac{31}{15}z-\frac{4}{15}z^2\bigg{\}}\\
&+C_A C_F\bigg{\{}\left(4+4z^{-1}\right)
\left[-\frac{1}{2}\widetilde{\Phi}(z)+\frac{1}{4}\ln^2 z-\frac{\zeta(2)}{2}\right]
-\frac{8}{z}{\rm Li}_2 (1-z)\\
&+\left(-2+2z^{-1}\right)\ln^2 (1-z)-\left(6+8z^{-1}\right)\ln^2 z+\left(18-\frac{58}{3}z^{-1}
+2z-\frac{2}{3}z^2\right)\ln(1-z)\\
&+\zeta(2)\left(-8+12z^{-1}\right)
+\left(14-\frac{44}{3}z^{-1}+4z-\frac{2}{3}z^2\right)\ln z-8 \ln z \ln (1-z) \\
&-\frac{40}{3}+\frac{56}{3}z^{-1}
-\frac{20}{3}z+\frac{4}{3}z^2\bigg{\}}\Bigg],
\end{split}
\eeq}
where the polylogarithms ${\rm Li}_n$ for $n=2,3$, the harmonic sum $S_{1,2}$ and the function
$\widetilde{\Phi}$ are defined as
\beq
\begin{split}
{\rm Li}_2 (x)=&-\int_0^x \frac{dy}{y}\ln(1-y),\\
{\rm Li}_3 (x)=&\int_0^x \frac{dy}{y}\ln \frac{y}{x} \ln(1-y),\\
S_{1,2}(x)=&\frac{1}{2}\int_0^x \frac{dy}{y}\ln^2(1-y),\\
\widetilde{\Phi}(x)=&\int_{x/(1+x)}^{1/(1+x)} \frac{dz}{z}
\ln \frac{1-z}{z}.
\end{split}
\eeq
To calculate the Mellin transform of the coefficient functions we require only the results
in Table \ref{Melltab1}, which are obtained from Ref.\ \cite{Blumlein:1998if}.
Formally, $\eta=(-1)^n$, although to analytically continue
the results in the right hand columns to complex $n$ requires taking
\beq
\eta=\exp\left[-\pi {\rm Im}(n)\right]
\left[\cos \left(\pi {\rm Re}(n)\right)+i\sin \left(\pi {\rm Re}(n)\right)\right].
\eeq
The harmonic sums $S_j (n)$ are defined for integer $n$ by
\beq
S_j(n)=\sum_{k=1}^n\frac{1}{k^j}.
\label{defofharmsumsinN}
\eeq
For complex $n$, the harmonic sums with $j=1,2,3$ can be calculated using the
results \cite{Abramowitz:1968}
\beq
\begin{split}
S_1(n)=&\psi(n+1)+\gamma_E,\\
S_2(n)=&-\psi'(n+1)+\zeta(2)\ {\rm and}\\
S_3(n)=&\frac{1}{2}\psi''(n+1)+\zeta(3),
\end{split}
\label{Snintermsofpsiderivs}
\eeq
where $\psi(n)$ and its derivatives can be evaluated for large $n$ using
{\small
\beq
\psi(n)=\ln n -\frac{1}{2n}-\frac{1}{12n^2}+\frac{1}{120 n^4}-\frac{1}{252 n^6}
+\frac{1}{240 n^8}-\frac{1}{132 n^{10}}+O\left(\frac{1}{n^{12}}\right).
\label{psiatlargeN}
\eeq}
As noted in Ref.\ \cite{Gluck:1991ee}, the harmonic sums when 
$n$ is small can be calculated by using Eq.\ (\ref{defofharmsumsinN})
to write $S_j(n)$ in the form
\beq
S_j(n)=S_j(n+r)-\sum_{k=1}^{r}\frac{1}{(k+n)^j},
\label{SnNandSnNprrel}
\eeq
where $r$ is chosen such that ${\rm Re}(n+r)$ is large enough to 
calculate $S_j(n+r)$ using Eq.\ (\ref{psiatlargeN}).

The function $\widetilde{S}(n)$ (also known as $S_{-2,1}(n)$) is defined for integer $n$ by
\beq
\widetilde{S}(n)=\sum_{k=1}^n \frac{(-1)^k}{k^2}\sum_{r=1}^k \frac{1}{r}.
\label{defoftildeSforintn}
\eeq
A method for calculating $\widetilde{S}(n)$ is given
in Ref.\ \cite{Gluck:1991ee}, and
generalized for other functions appearing in perturbative QCD up to two
loop order in Ref.\ \cite{Blumlein:2000hw}. 
Here we present an alternative method which is similar
to the calculation of the $S_j(n)$ using Eqs.\ (\ref{psiatlargeN}) and (\ref{SnNandSnNprrel}).
Firstly, we analytically continue Eq.\ (\ref{defoftildeSforintn})
to complex values of $n$ by writing it in the form
\beq
\widetilde{S}(n)=\sum_{k=1}^{\infty} \frac{(-1)^k}{k^2}\sum_{r=1}^k \frac{1}{r}
-(-1)^n\sum_{k=1}^{\infty} \frac{(-1)^k}{(k+n)^2} S_1(k+n).
\label{defoftildeSforcompn}
\eeq
The first term gives $-\frac{5}{8}\zeta(3)$. 
For all values of $n$ except for $n=-1,-2,...$, the second term converges, but very slowly.
Instead, we use Eq.\ (\ref{defoftildeSforintn}) to write $\widetilde{S}(n)$
in the form
\beq
\widetilde{S}(n)=\widetilde{S}(n+r)-(-1)^n\sum_{k=1}^r \frac{(-1)^k}{(k+n)^2}S_1(n+k),
\eeq
where $r$ is chosen such that ${\rm Re}(n+r)$ is large, and calculate $\widetilde{S}(n+r)$
as a series in $1/(n+r)$. For this purpose, we write $S_1(n+k)$ in the form
\beq
S_1(n+k)=\ln n +\gamma_E +\sum_{l=1}^{\infty}\frac{1}{n^l}\sum_{m=0}^l A_m^l k^m,
\eeq
where the $A_m^l$ may be easily calculated using the first relation in Eq.\ (\ref{Snintermsofpsiderivs})
and Eq.\ (\ref{psiatlargeN}), then we expand the second term in Eq.\ (\ref{defoftildeSforcompn}) 
in $\frac{1}{n}$, making use of the relation
\beq
\sum_{k=1}^{\infty} (-1)^k k^r=\left(x\frac{d}{dx}\right)^r\left(\frac{1}{1+x}-1\right)\Bigg{|}_{x=1},
\eeq
to obtain the result
\beq
\begin{split}
\sum_{k=1}^{\infty} &\frac{(-1)^k}{(k+n)^2}\sum_{r=1}^{k+n} \frac{1}{r}\\
&=(\ln n+\gamma_E)\sum_{p=0}^{\infty} \frac{1}{n^{p+2}}(-1)^p (p+1)
\left(x\frac{d}{dx}\right)^p\left(\frac{1}{1+x}-1\right)\Bigg{|}_{x=1}\\
&+\sum_{p=1}^{\infty}\frac{1}{n^{p+2}}\sum_{r=0}^{p-1}(-1)^r (r+1)\sum_{m=0}^{p-r}
A_m^{p-r}\left(x\frac{d}{dx}\right)^{r+m}\left(\frac{1}{1+x}-1\right)\Bigg{|}_{x=1}.
\end{split}
\eeq
The coefficient of the terms $1/n^p$ may now be evaluated, and we find
\beq
\begin{split}
\widetilde{S}&(n)=-\frac{5}{8}\zeta(3)\\
&-\eta(n)\Bigg[\left(\ln n +\gamma_E \right)\left(-\frac{1}{2n^2}
+\frac{1}{2n^3}-\frac{1}{2n^5}+\frac{3}{2n^7}-\frac{17}{2n^9}+O\left(\frac{1}{n^{11}}\right)\right)\\
&-\frac{1}{2n^3}+\frac{5}{12n^4}+\frac{11}{24n^5}-\frac{151}{240n^6}-\frac{469}{240n^7}
+\frac{331}{126n^8}+\frac{67379}{5040n^9}+O\left(\frac{1}{n^{10}}\right)\Bigg].
\end{split}
\eeq

Since all occurrences of $\eta^2$ must be replaced by unity in the analytic continuation,
in the second row of Table \ref{Melltab1},
we have made the necessary adjustments to that result presented in Ref.\ \cite{Blumlein:1998if}.

FORTRAN routines for the longitudinal coefficient functions to NLO in Mellin
space are provided at \verb+http://www.desy.de/~simon/cf_long.html+.

\begin{acknowledgements}

The authors would like to thank M.\ Heinz for providing them with the
numerical values for the $p+p \rightarrow K^0_S +X$ STAR data shown graphically in
Ref.\ \cite{MH}. This work was supported in part by the Deutsche Forschungsgemeinschaft through
Grant No.\ KN~365/3-1, and by the Bundesministerium f\"ur Bildung und Forschung
through Grant No.\ 05~HT4GUA/4.

\end{acknowledgements}

\renewcommand{\thetable}{\arabic{table}}

\newpage
\begin{table}[h]
\begin{small}
\renewcommand{\arraystretch}{1.1}
\caption{\label{pars}Values and errors of $N$, $\alpha$ and $\beta$ in
Eq.\ (\ref{param}) resulting from the fit.}
\begin{center}
\begin{tabular}{c|l|ccc}
\hline \hline 
Hadron & Flavour & $N$ & $\alpha$ & $\beta$ \\
\hline 
$\pi^\pm$ & $d$  & 0.833 $\pm$ 0.012 & $-$1.17 $\pm$ 0.01  & 1.39 $\pm$ 0.02 \\
      & $u$    & 0.447 $\pm$ 0.007   & $-$1.58 $\pm$ 0.01  & 1.01 $\pm$ 0.02 \\
      & $s$    & 0.519 $\pm$ 0.035   & $-$0.365 $\pm$ 0.066 & 1.96 $\pm$ 0.10 \\
      & $c$    & 1.56 $\pm$ 0.03    & $-$1.03 $\pm$ 0.01   & 3.58 $\pm$ 0.07 \\
      & $b$    & 0.139 $\pm$ 0.001       & $-$2.24 $\pm$ 0.01   & 2.77 $\pm$ 0.05 \\
      & $g$    & 429 $\pm$ 3         & 2.00 $\pm$ 0.01      & 5.82 $\pm$ 0.01 \\ \hline
$K^\pm$   & $d$  & 2245 $\pm$ 465  & 4.14 $\pm$ 0.18      & 12.0 $\pm$ 0.5 \\ 
      & $u$    & 10.9 $\pm$ 0.7    & 1.72 $\pm$ 0.08      & 3.44 $\pm$ 0.08 \\
      & $s$    & 0.529 $\pm$ 0.012 & $-$0.787 $\pm$ 0.027 & 0.915 $\pm$ 0.027 \\
      & $c$    & 2.28 $\pm$ 0.09   & $-$0.488 $\pm$ 0.028 & 3.79 $\pm$ 0.09 \\
      & $b$    & 1.13 $\pm$ 0.03   & $-$0.960 $\pm$ 0.016 & 6.22 $\pm$ 0.09 \\
      & $g$    & 15.9 $\pm$ 0.5    & 2.72 $\pm$ 0.05      & 2.45 $\pm$ 0.03 \\ \hline
$p/\overline{p}$   & $d$ & 146 $\pm$ 22 & 2.30 $\pm$ 0.12    & 10.4 $\pm$ 0.4 \\ 
      & $u$    & 0.0182 $\pm$ 0.0014    & $-$2.37 $\pm$ 0.05 & 0.507 $\pm$ 0.125 \\
      & $s$    & 1859 $\pm$ 648         & 6.67 $\pm$ 0.49    & 9.17 $\pm$ 0.53 \\
      & $c$    & 12.0 $\pm$ 1.4         & 0.860 $\pm$ 0.089  & 7.50 $\pm$ 0.28 \\
      & $b$    & 1571 $\pm$ 103         & 2.19 $\pm$ 0.04    & 19.0 $\pm$ 0.3 \\
      & $g$    & 0.867 $\pm$ 0.023      & 1.13 $\pm$ 0.06    & 0.854 $\pm$ 0.020 \\
\hline \hline
\end{tabular}
\end{center}
\end{small}
\end{table}

\begin{table}[h]
\begin{footnotesize}
\renewcommand{\arraystretch}{1.1}
\caption{$\chi^2_{\mathrm{DF}}$ values
obtained from the measured light quark tagging probabilities $\eta_a^h$
at $\sqrt{s}=91.2$ GeV in Ref.\ \cite{Abbiendi:1999ry}.\label{chiDFforlqetas}}
\begin{center}
\begin{tabular}{||c||l|l|l||}
\hline\hline
\backslashbox{$a$}{$h$} & $\pi^{\pm}$ & $K^{\pm}$ & $p/\overline{p}$ \\
\hline\hline
$d$ & 5.05 & 0.47 & 2.16 \\
\hline
$u$ & 4.87 & 0.43 & 1.20 \\
\hline
$s$ & 2.69 & 0.92 & 1.23 \\
\hline\hline
\end{tabular}
\end{center}
\end{footnotesize}
\end{table}

\begin{table}[h]
\begin{footnotesize}
\renewcommand{\arraystretch}{1.1}
\caption{As in Table \ref{chiDFforlqetas}, but for the heavy quarks.
\label{chiDFforhqetas}}
\begin{center}
\begin{tabular}{||c||l|l|l||}
\hline\hline
\backslashbox{$a$}{$h$} & $\pi^{\pm}$ & $K^{\pm}$ & $p/\overline{p}$ \\
\hline\hline
$b$ & 17.9 & 10.9 & 7.64 \\
\hline
$c$ & 24.1 & 11.8 & 2.96 \\
\hline\hline
\end{tabular}
\end{center}
\end{footnotesize}
\end{table}

\begin{table}[ht]
\begin{footnotesize}
\renewcommand{\arraystretch}{1.1}
\caption{CM energies, types of data, and $\chi^2_{\mathrm{DF}}$ values
for various data samples.
Samples not used in the fits are marked by asterisks. ($\{h\}$ refers to a sum
over light charged hadrons and $\{q\}$ refers to a sum over all 5 flavours of quarks.)
\label{allchis}}
\begin{center}
\begin{tabular}{c|l|lll}
\hline\hline
 $\sqrt{s}$ [GeV] & Data type & 
\multicolumn{3}{c}{\makebox[5cm][c]{$\chi^2_{\mathrm{DF}}$}} \\
\hline
29.0 & $F^{\{h\}}_{uds}$ & 3.44 \cite{Aihara:1986mv}$^*$ \\
     & $F^{\{h\}}_c$ & 2.56 \cite{Aihara:1986mv}$^*$ \\
     & $F^{\{h\}}_b$ & 1.74 \cite{Aihara:1986mv}$^*$ \\
\hline
     & $F^\pi_{\{q\}}$ & 0.80 \cite{Aihara:1988fc} \\ 
     & $F^\pi_{uds}$ & 1.01 \cite{Aihara:1986mv} \\
     & $F^\pi_c$ & 2.51 \cite{Aihara:1986mv} \\
     & $F^\pi_b$ & 2.14 \cite{Aihara:1986mv} \\
\hline
     & $F^K_{\{q\}}$ & 0.37 \cite{Aihara:1988fc} \\ 
\hline
     & $F^p_{\{q\}}$ & 0.80 \cite{Aihara:1988fc}\\
\hline
91.2 & $F^{\{h\}}_{\{q\}}$ &
2.61 \cite{Abreu:1998vq}$^*$ & 105 \cite{PadillaAranda:1995wi}$^*$ & 22.0 \cite{Ackerstaff:1998hz}$^*$ \\
 & & 4.99 \cite{Abe:1998zs}$^*$ \\
 & $F^{\{h\}}_{\{uds\}}$ &
1.10 \cite{Abreu:1998vq}$^*$ & 64.8 \cite{Buskulic:1995aw}$^*$ & 2.08 \cite{Ackerstaff:1998hz}$^*$ \\
 & $F^{\{h\}}_c$ &
 & 34.4 \cite{Buskulic:1995aw}$^*$ & 0.57 \cite{Ackerstaff:1998hz}$^*$ \\
 & $F^{\{h\}}_b$ &
0.21 \cite{Abreu:1998vq}$^*$ & 183.6 \cite{Buskulic:1995aw}$^*$ & 5.90 \cite{Ackerstaff:1998hz}$^*$ \\
\hline
 & $F^\pi_{\{q\}}$ &
0.98 \cite{Buskulic:1994ft} & 1.13 \cite{Abreu:1998vq} & 1.82 \cite{Abe:1998zs} \\
 & $F^\pi_{\{uds\}}$ &
 & 1.82 \cite{Abreu:1998vq} & 1.12 \cite{Abe:1998zs} \\
 & $F^\pi_c$ &
 & & 1.08 \cite{Abe:1998zs} \\
 & $F^\pi_b$ &
 & 0.40 \cite{Abreu:1998vq} & 0.67 \cite{Abe:1998zs} \\
\hline
& $F^K_{\{q\}}$ &
0.52 \cite{Buskulic:1994ft} & 0.31 \cite{Abreu:1998vq} & 0.52 \cite{Abe:1998zs} \\
& $F^K_{\{uds\}}$ &
 & 0.31 \cite{Abreu:1998vq} & 0.83 \cite{Abe:1998zs} \\
 & $F^K_c$ &
 & & 1.79 \cite{Abe:1998zs} \\
 & $F^K_b$ &
 & 0.10 \cite{Abreu:1998vq} & 1.17 \cite{Abe:1998zs} \\
\hline
 & $F^p_{\{q\}}$ &
0.65 \cite{Buskulic:1994ft} & 0.11 \cite{Abreu:1998vq} & 0.69 \cite{Abe:1998zs} \\
 & $F^p_{\{uds\}}$ &
 & 0.22 \cite{Abreu:1998vq} & 1.43 \cite{Abe:1998zs} \\
 & $F^p_c$ &
 & & 0.74 \cite{Abe:1998zs} \\
 & $F^p_b$ &
 & 0.52 \cite{Abreu:1998vq} & 1.14 \cite{Abe:1998zs} \\
\hline
$E_{\mathrm{jet}}$ [GeV] \\
\hline
26.2 & $D_g^{\{h\}}$ & 23.6 \cite{Barate:1998cp}$^*$ \\
40.1 & $D_g^{\{h\}}$ & 4.36 \cite{Abbiendi:1999pi}$^*$ \\
\hline \hline
\end{tabular}
\end{center}
\end{footnotesize}
\end{table}

\renewcommand{\thefigure}{\arabic{figure}}

\newpage
\begin{figure}[ht]
\centering
\setlength{\epsfxsize}{13.5cm}
\begin{minipage}[ht]{\epsfxsize}
\centerline{\mbox{\epsffile{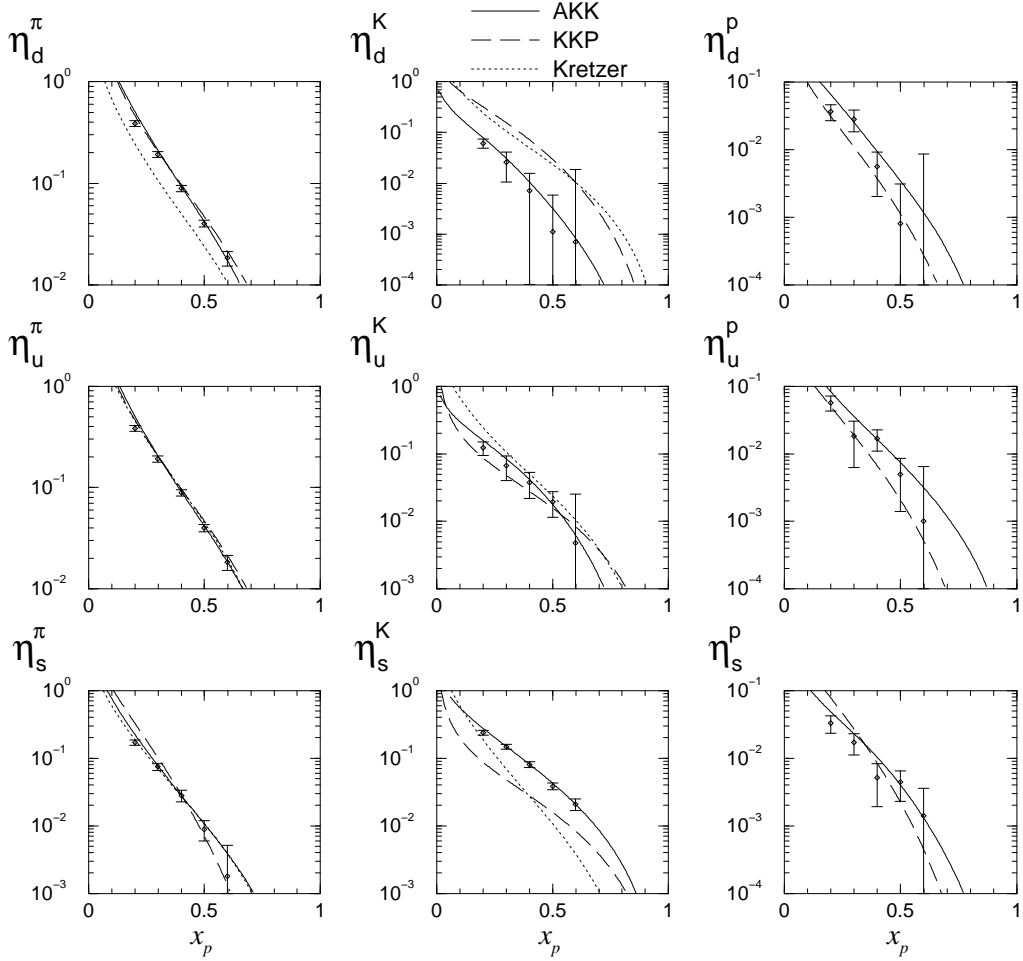}}}
\end{minipage}
\caption{
Light quark probabilities $\eta^h_a(x_p,s)$ at $\sqrt{s}=91.2$ GeV. The dashed curves
are calculated using the FFs obtained in Ref.\ \cite{Kniehl:2000fe}, 
the dotted curves are calculated from the $(x,M_f^2)$ grid of FFs obtained
from the analysis of Ref.\ \cite{Kretzer:2000yf} (in which no $p/\overline{p}$ FFs
are obtained), and the solid curves are calculated
using the FFs obtained in the analysis of this paper. The corresponding measured OPAL probabilites of
Ref.\ \cite{Abbiendi:1999ry} are also shown.
\label{fig1}}
\end{figure}

\begin{figure}[ht]
\centering
\setlength{\epsfxsize}{13.5cm}
\begin{minipage}[ht]{\epsfxsize}
\centerline{\mbox{\epsffile{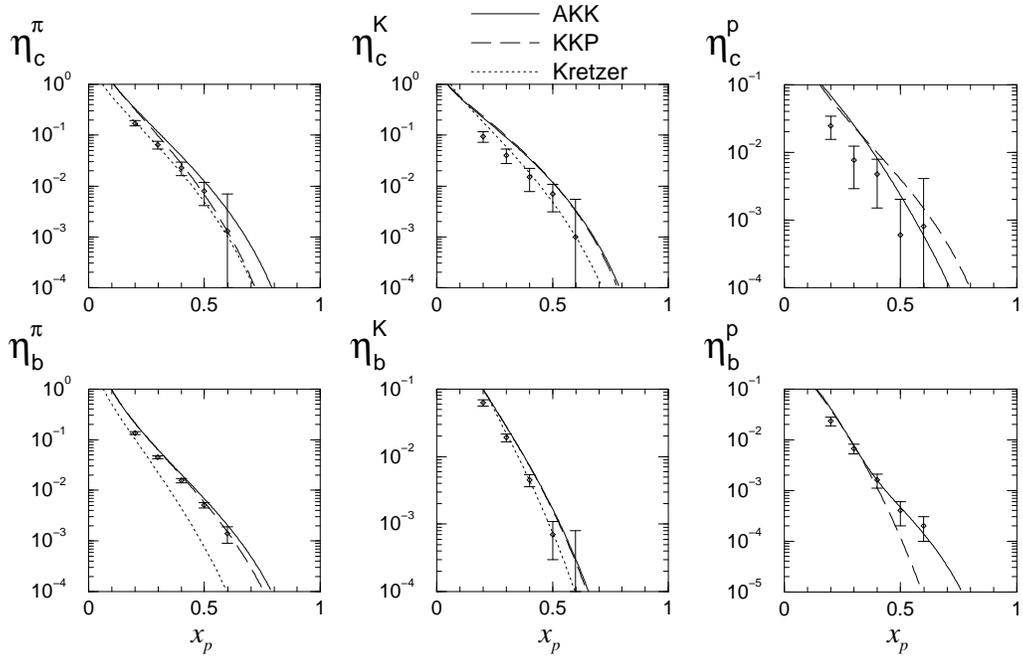}}}
\end{minipage}
\caption{
As in Fig.\ \ref{fig1}, but for the heavy quark probabilities.
\label{fig7}}
\end{figure}

\begin{figure}[hb]
\centering
\setlength{\epsfxsize}{13.5cm}
\begin{minipage}[ht]{\epsfxsize}
\centerline{\mbox{\epsffile{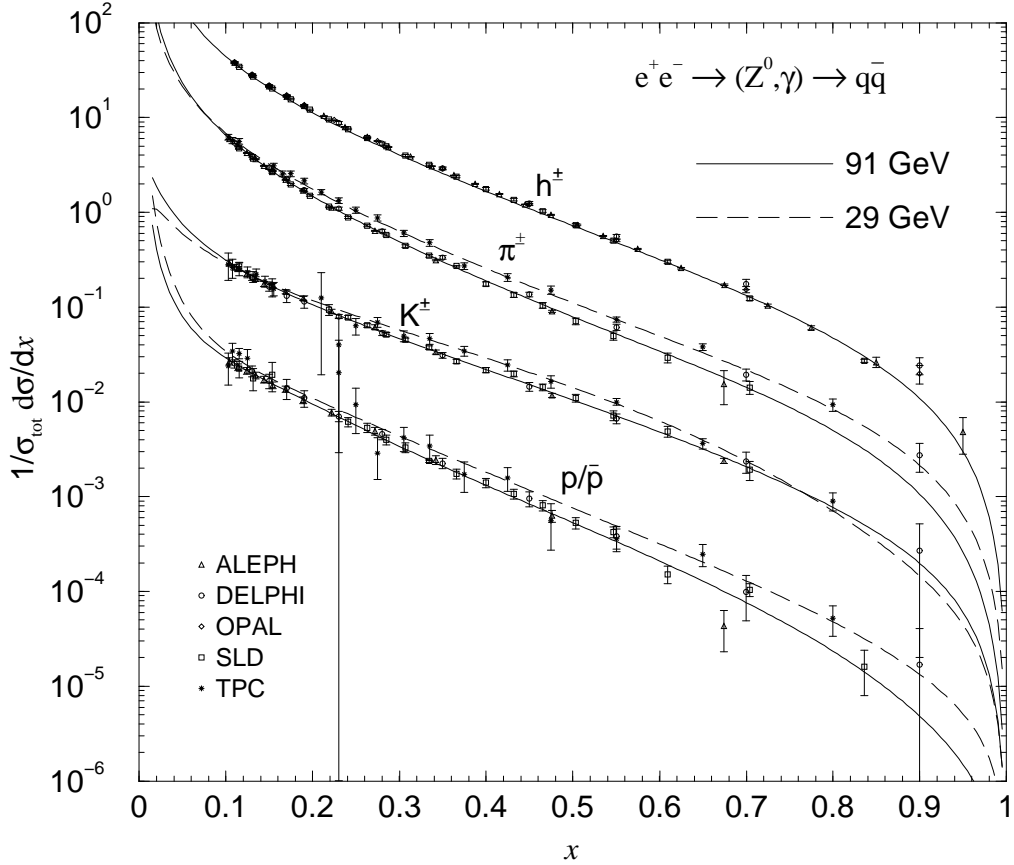}}}
\end{minipage}
\caption{
Normalized differential cross section of inclusive hadron production.
The curves are calculated from the FFs obtained in our analysis, at 29 (dashed line) and
91.2 (solid line) GeV.
The upmost, second, third and lowest curves refer to charged hadrons, $\pi^{\pm}$,
$K^{\pm}$ and $p/\overline{p}$ respectively. 
The differential cross section for the charged hadron curve was calculated by taking the sum of the 
differential cross sections for the three lightest charged hadrons.
The ALEPH \cite{Buskulic:1994ft}, DELPHI \cite{Abreu:1998vq}, OPAL \cite{Ackerstaff:1998hz},
SLD \cite{Abe:1998zs} and TPC \cite{Aihara:1988fc} data sets are shown.
The charged hadron data are shown just for comparison, but were not used in the fit.
Each curve or pair of curves and the corresponding data
is rescaled relative to the nearest upper one by a factor of 1/5.
\label{fig2}}
\end{figure}

\begin{figure}[ht]
\centering
\setlength{\epsfxsize}{13.5cm}
\begin{minipage}[ht]{\epsfxsize}
\centerline{\mbox{\epsffile{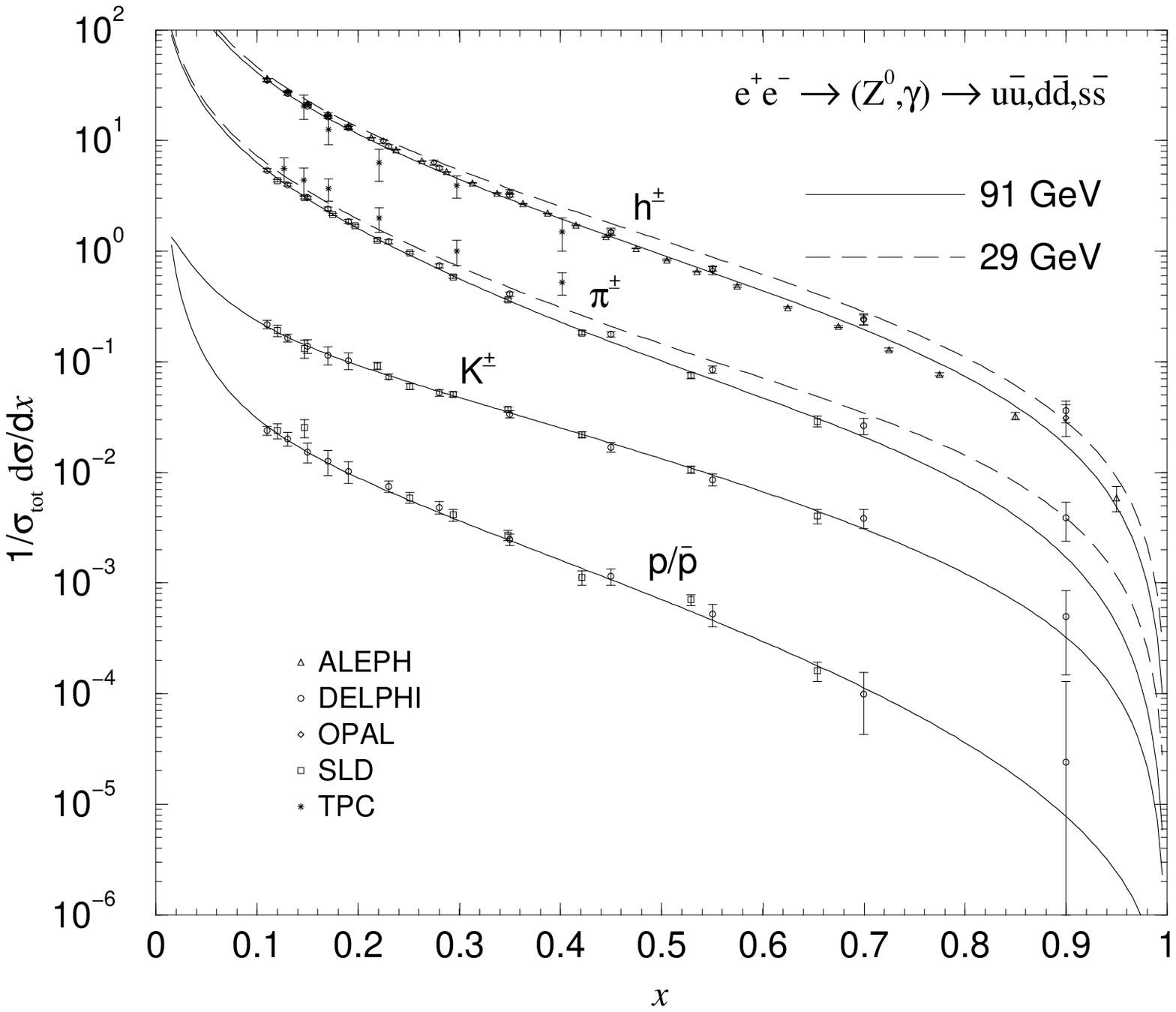}}}
\end{minipage}
\caption{
As in Fig.\ \ref{fig2}, but for the light quark tagged cross sections.
The ALEPH \cite{Buskulic:1994ft}, DELPHI \cite{Abreu:1998vq}, OPAL \cite{Ackerstaff:1998hz},
SLD \cite{Abe:1998zs} and TPC \cite{Aihara:1986mv} data sets are shown.
The charged hadron data are shown just for comparison, but were not used in the fit.
\label{fig3}}
\end{figure}

\begin{figure}[ht]
\centering
\setlength{\epsfxsize}{13.5cm}
\begin{minipage}[ht]{\epsfxsize}
\centerline{\mbox{\epsffile{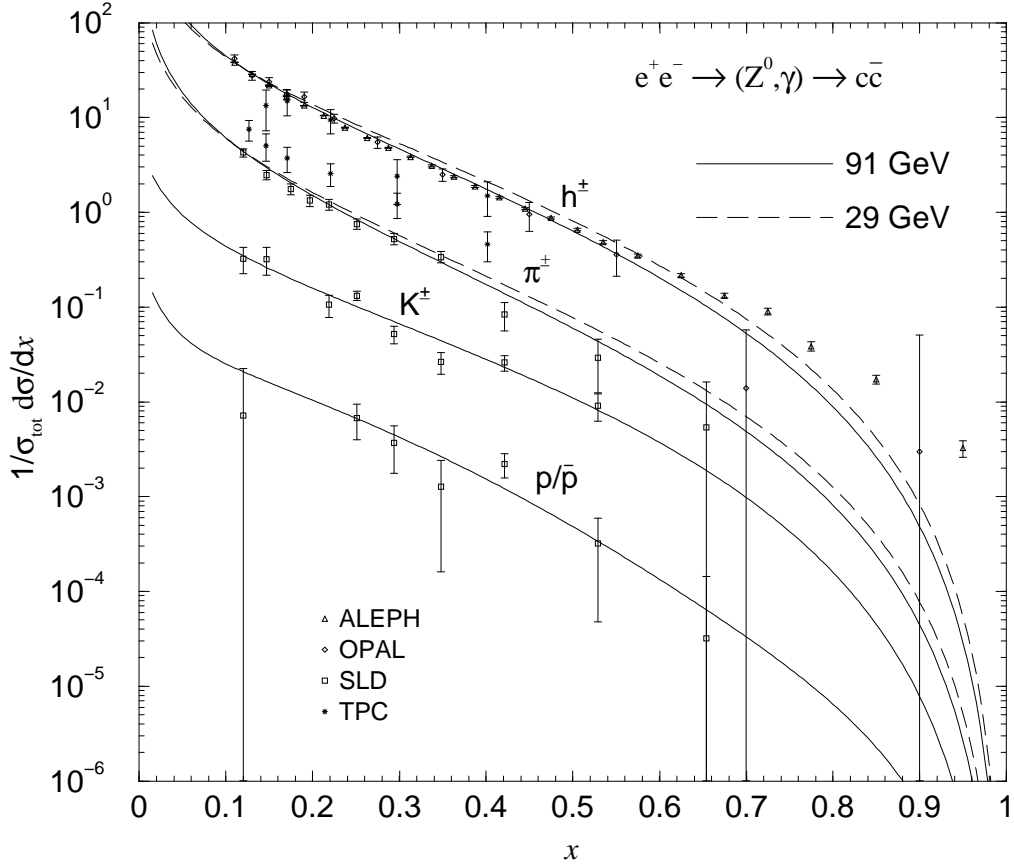}}}
\end{minipage}
\caption{
As in Fig.\ \ref{fig2}, but for the $c$ quark tagged cross sections.
The ALEPH \cite{Buskulic:1994ft}, OPAL \cite{Ackerstaff:1998hz} ,
SLD \cite{Abe:1998zs} and TPC \cite{Aihara:1986mv} data sets are shown. 
The two SLD data points at $x=0.654$
are for the pion (upper) and proton (lower).
The charged hadron data are shown just for comparison, but were not used in the fit.
\label{fig4}}
\end{figure}

\begin{figure}[ht]
\centering
\setlength{\epsfxsize}{13.5cm}
\begin{minipage}[ht]{\epsfxsize}
\centerline{\mbox{\epsffile{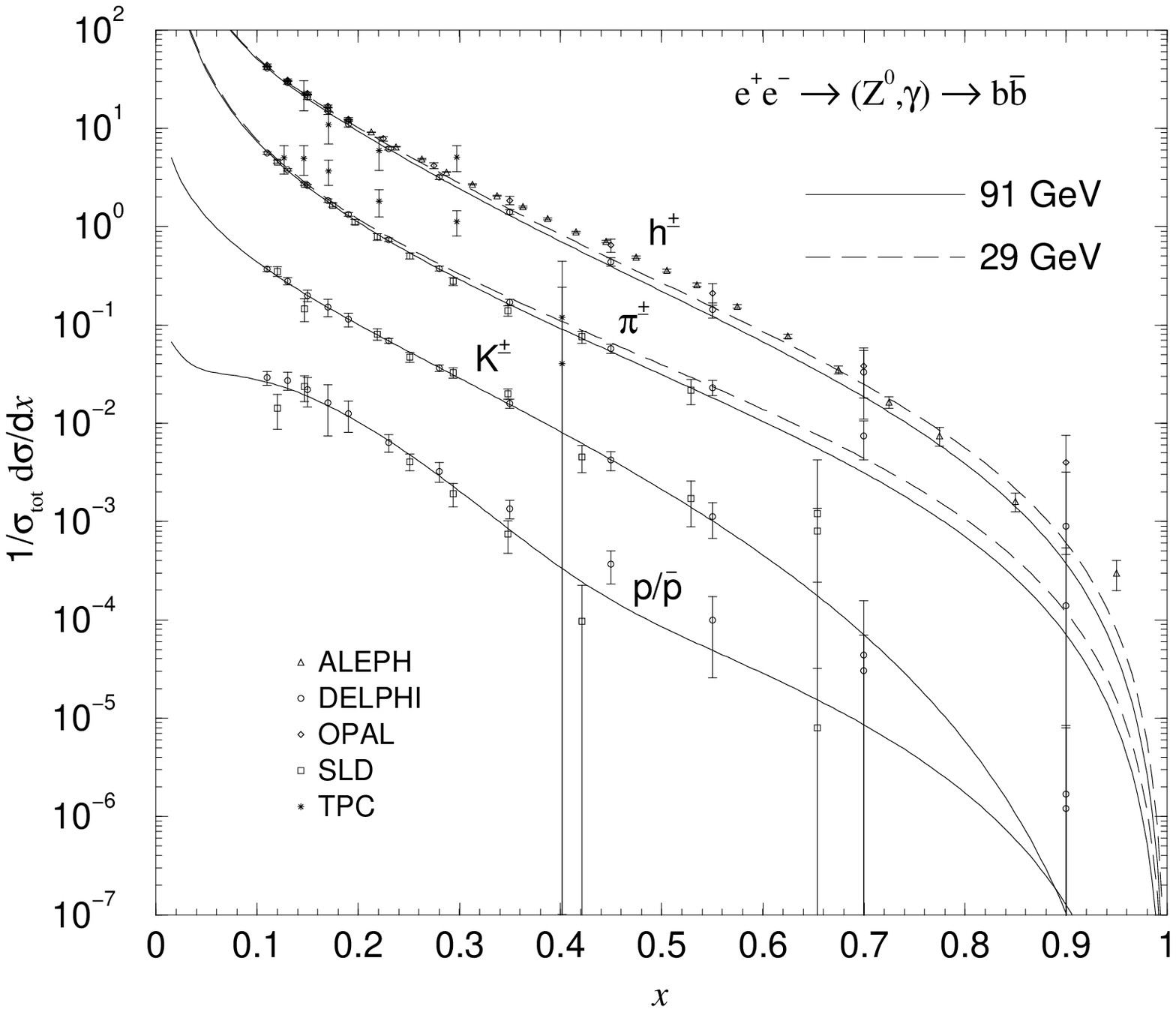}}}
\end{minipage}
\caption{
As in Fig.\ \ref{fig2}, but for the $b$ quark tagged cross sections.
The ALEPH \cite{Buskulic:1994ft}, DELPHI \cite{Abreu:1998vq}, OPAL \cite{Ackerstaff:1998hz},
SLD \cite{Abe:1998zs} and TPC \cite{Aihara:1986mv} data sets are shown.
The charged hadron data are shown just for comparison, but were not used in the fit.
\label{fig5}}
\end{figure}

\begin{figure}[ht]
\centering
\setlength{\epsfxsize}{13.5cm}
\begin{minipage}[ht]{\epsfxsize}
\centerline{\mbox{\epsffile{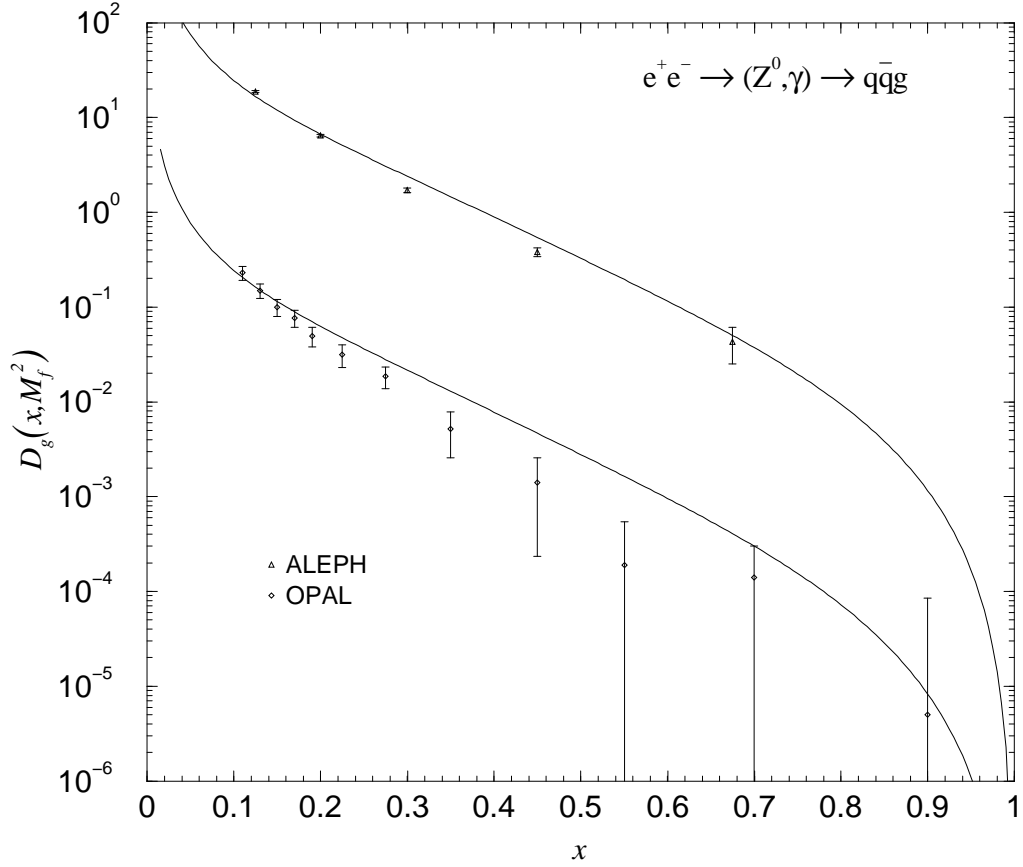}}}
\end{minipage}
\caption{
Gluon FF for charged-hadron production at $M_f=52.4$ and $80.2$ GeV.
The curves are calculated from the FFs obtained in our analysis.
The three-jet data from ALEPH \cite{Barate:1998cp}, with $E_{\rm jet}=26.2$ GeV, 
and from OPAL \cite{Abbiendi:1999pi},
with $E_{\rm jet}=40.1$ GeV, are shown. The OPAL data and its corresponding
curve are rescaled by a factor of 1/100.
\label{fig6}}
\end{figure}

\begin{figure}[ht]
\centering
\setlength{\epsfxsize}{13.5cm}
\begin{minipage}[ht]{\epsfxsize}
\centerline{\mbox{\epsffile{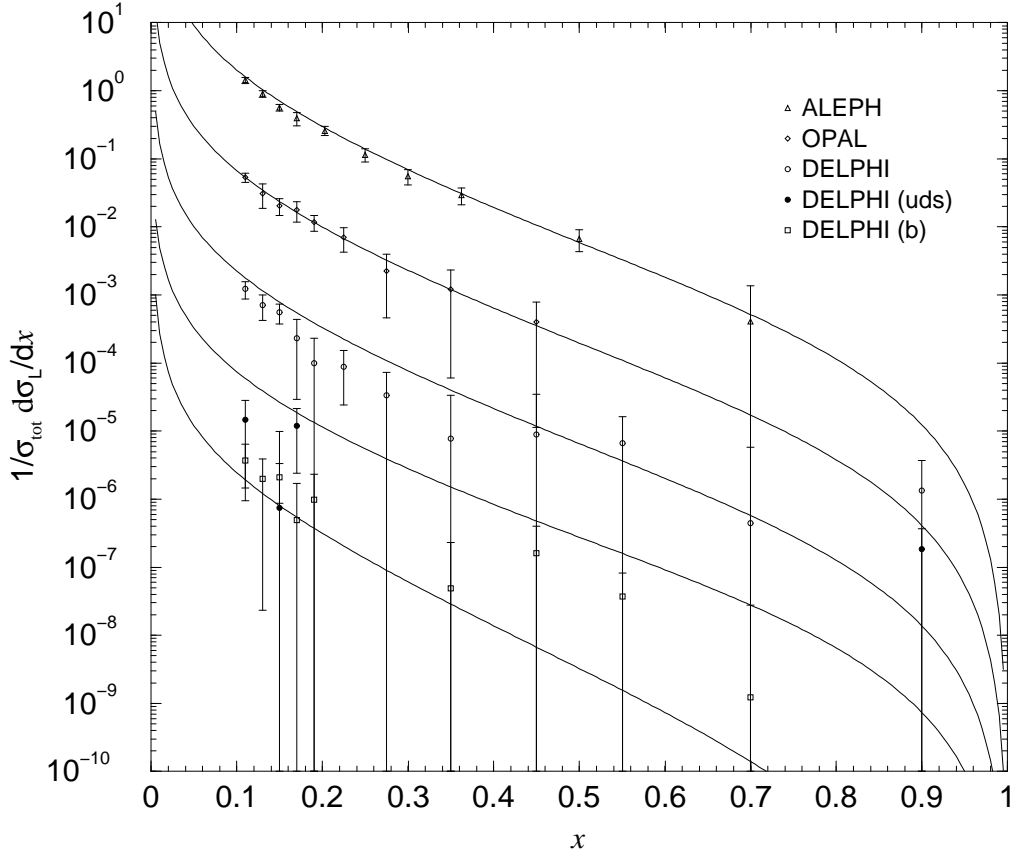}}}
\end{minipage}
\caption{Normalized longitudinal differential cross section of inclusive charged 
hadron production at $\sqrt{s}=91.2$ GeV.
The curves are calculated from the FFs obtained in our analysis.
The data sets shown are from ALEPH \cite{Buskulic:1995aw}, OPAL \cite{Akers:1995wt} 
and DELPHI \cite{Abreu:1997ir} without flavour separation and DELPHI \cite{Abreu:1997ir} 
for light and $b$ quark tagged cross sections.
Each curve is rescaled relative to the nearest upper one by a factor of 1/30.
\label{fig8}}
\end{figure}

\begin{figure}[ht]
\centering
\setlength{\epsfxsize}{13.5cm}
\begin{minipage}[ht]{\epsfxsize}
\centerline{\mbox{\epsffile{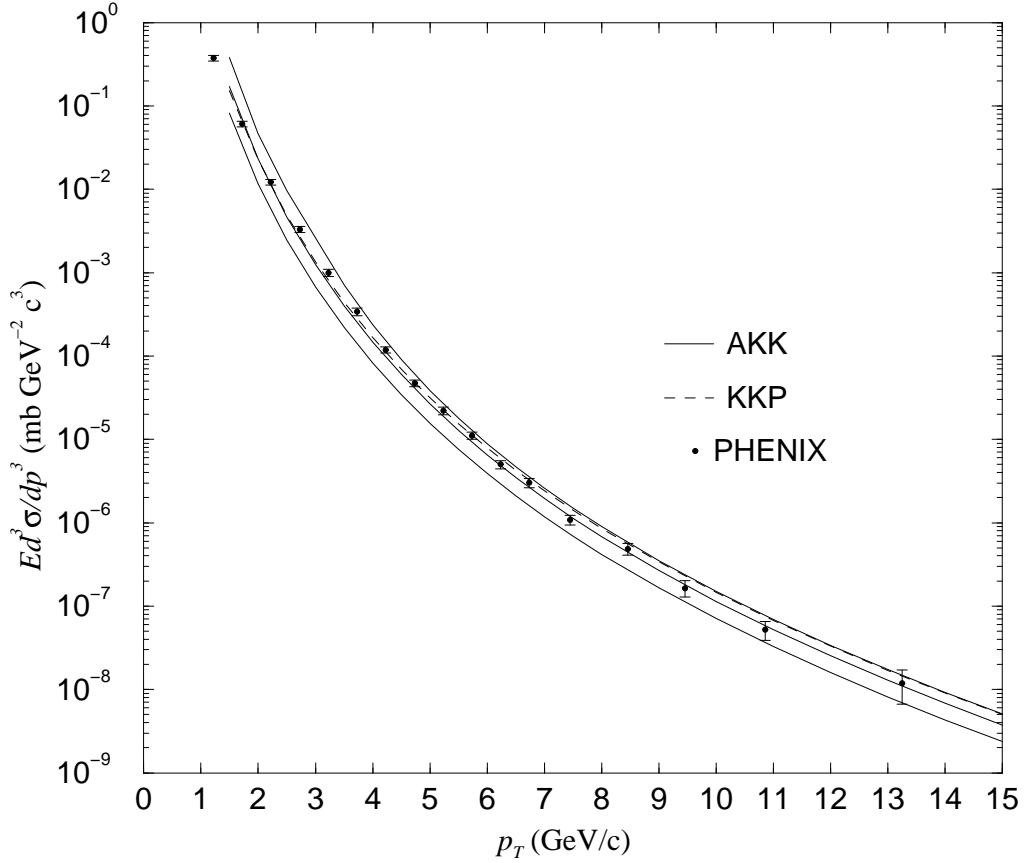}}}
\end{minipage}
\caption{The invariant differential cross section for inclusive $\pi^0$ production 
in $p+p$ collisions at $\sqrt{s}=200$ GeV.
Data from the PHENIX Collaboration \cite{Adler:2003pb} are shown, without the absolute
9.6\% normalization error. Compared with this data are the
cross sections calculated from the FFs obtained in this paper (labelled AKK)
and that from the FFs of Ref.\ \cite{Kniehl:2000fe} (labelled KKP).
The upper, central and lower AKK curves are calculated with
$k=1/4$, 1 and 4 respectively.
\label{fig9}}
\end{figure}

\begin{figure}[ht]
\centering
\setlength{\epsfxsize}{13.5cm}
\begin{minipage}[ht]{\epsfxsize}
\centerline{\mbox{\epsffile{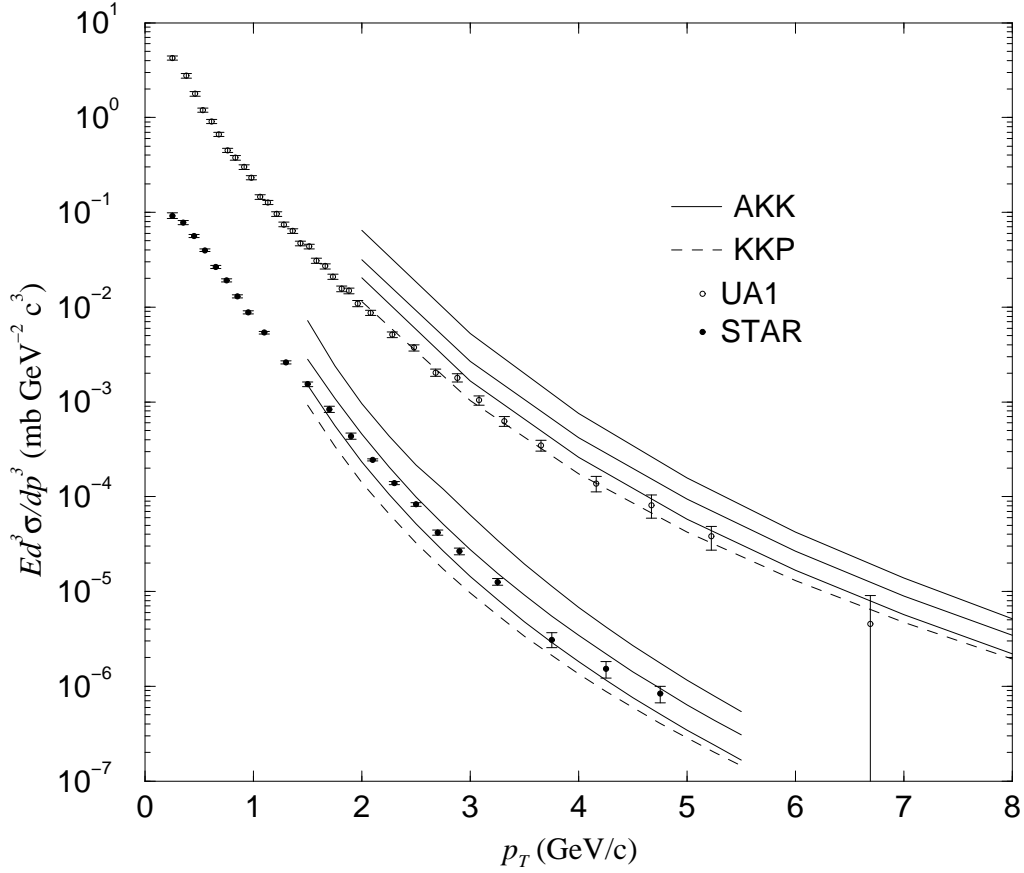}}}
\end{minipage}
\caption{As in Fig.\ \ref{fig9}, but for the invariant differential cross section 
for inclusive $K^0_S$ production in $pp$ 
collisions at $\sqrt{s}=200$ GeV compared with data from the STAR Collaboration
\cite{MH}, and in
$p+\overline{p}$ collisions at $\sqrt{s}=630$ GeV compared 
with data from the UA1 Collaboration \cite{Bocquet:1995jq}. For clarity, the former
results have been divided by a factor of 30.
\label{fig10}}
\end{figure}

\renewcommand{\thetable}{A}

\begin{table}[hhhh]
\noindent
\caption{Mellin transforms required for the Mellin space conversion of the NLO coefficient
functions of the longitudinal cross section. \label{Melltab1}}
\begin{center}
\vspace{2mm}
\begin{tabular}{||c|l||}
\hline
\hline
&   \\[-3mm]
\multicolumn{1}{||c|}{${\ds f(z)}$}& \multicolumn{1}{c||}{${\ds \int_0^1 dz z^{n-1} f(z)}$} \\
&   \\[-3mm]
\hline
\hline
&  \\[-3mm]
${\ds z^r }$ & $\frac{1}{n+r}$\\
&  \\[-3mm]
\hline
& \\[-3mm]
${\ds z^r \ln z }$ & $-\frac{1}{(n+r)^2}$\\
&  \\[-3mm]
\hline
& \\[-3mm]
${\ds z^r \ln^2 z }$ & $\frac{2}{(n+r)^3}$\\
&  \\[-3mm]
\hline
& \\[-3mm]
${\ds z^r \ln (1-z) }$ & $-\frac{S_1(n+r)}{n+r}$\\
&  \\[-3mm]
\hline
& \\[-3mm]
${\ds z^r \ln^2 (1-z) }$ & $\frac{S_1^2 (n+r)+S_2(n+r)}{n+r}$\\
&  \\[-3mm]
\hline
& \\[-3mm]
${\ds \ln z \ln (1-z) }$ & $\frac{S_1(n)}{n^2}+\frac{1}{n}\left[S_2(n)-\zeta(2)\right]$\\
&  \\[-3mm]
\hline
& \\[-3mm]
${\ds z^r {\rm Li}_2(1-z) }$ & $-\frac{1}{n+r}\left[S_2(n+r)-\zeta(2)\right]$\\
&  \\[-3mm]
\hline
& \\[-3mm]
${\ds S_{1,2}(1-z) }$ & $-\frac{1}{n}\left[S_3(n)-\zeta(3)\right]$\\
& \\[-3mm]
\hline
& \\[-3mm]
${\ds \ln (1+z)  }$ & $\frac{1}{n}\left[\eta(n) S_1(n)+\frac{1-\eta(n) }{2}
S_1\left(\frac{n-1}{2}\right)-\frac{1+\eta(n)}{2}S_1\left(\frac{n}{2}\right)\right]$\\
&$+[1-\eta(n) ]\frac{\ln 2}{n}$\\
& \\[-3mm]
\hline
& \\[-3mm]
${\ds {\rm Li}_3(-z) }$ & $-\eta(n)\frac{S_1(n)}{n^3}
-\frac{1-\eta(n) }{2n^3}\left[S_1\left(
\frac{n-1}{2}\right)+ 2\ln 2\right]$\\
&$+\frac{1+\eta(n)}{2n^3}S_1\left(\frac{n}{2}
\right)+\frac{\zeta(2)}{2n^2}-\frac{3\zeta(3)}{4n}$\\
&  \\
\hline
\hline
\end{tabular}
\end{center}
\end{table}

\renewcommand{\thetable}{A continued}

\begin{table}[hhhh]
\caption{}
\noindent
\begin{center}
\vspace{2mm}
\begin{tabular}{||c|l||}
\hline
\hline
&   \\[-3mm]
\multicolumn{1}{||c|}{${\ds f(z)}$}& \multicolumn{1}{c||}{${\ds \int_0^1 dz z^{n-1} f(z)}$} \\
&   \\[-3mm]
\hline
\hline
& \\[-3mm]
${\ds \ln z {\rm Li}_2(-z) }$ & $\frac{-\eta(n)}{n^2}
\left[\frac{2S_1(n)}{n}+S_2(n)\right]$\\
&$-\frac{1-\eta(n) }{2n^2}\left[\frac{2}{n}
S_1\left(\frac{n-1}{2}\right)+\frac{1}{2}S_2\left(
\frac{n-1}{2}\right)+\frac{4\ln 2}{n}\right]$\\
&$+\frac{1+\eta(n)}{2n^2}\left[\frac{2}{n}
S_1\left(\frac{n}{2}\right)+\frac{1}{2}S_2\left(\frac{n}{2}\right)
+\zeta(2)\right]$\\
&  \\[-3mm]
\hline
& \\[-3mm]
${\ds \ln^2 z\ln (1+z) }$ & $2\frac{\eta(n)}{n}\left[\frac{S_1(n)}{n^2}
+\frac{S_2(n)}{n}+S_3(n) -\frac{\zeta(2)}{2n}-\frac{3\zeta(3)}{4}\right]$\\
&$+\frac{1-\eta(n) }{2n}\left[\frac{2}{n^2}
S_1\left(\frac{n-1}{2}\right)+\frac{1}{n}S_2\left(
\frac{n-1}{2}\right)+\frac{1}{2}S_3\left(\frac{n-1}{2}\right)
+\frac{4\ln 2}{n^2}\right]$\\
& $-\frac{1+\eta(n)}{2n}\left[\frac{2}{n^2}
S_1\left(\frac{n}{2}\right) +\frac{1}{n}S_2\left(\frac{n}{2}\right)
+\frac{1}{2}S_3\left(\frac{n}{2}\right)\right]$\\
& \\
\hline
& \\[-3mm]
$ z^r\left[-\frac{1}{2}\widetilde{\Phi}(z)+\frac{1}{4}\ln^2 z-\frac{\zeta(2)}{2}\right] $ & 
$-\frac{1}{2}\bigg{\{}
\frac{1}{(n+r)^3}+2\frac{\eta(n) (-1)^r}{n+r}\left[S_2(n+r)
-\zeta(2)\right]$\\
&$-\frac{1+\eta(n) (-1)^r}{2(n+r)}\left[S_2\left(\frac{n+r}{2}\right)
-\zeta(2)\right]$\\
&$+\frac{1-\eta(n) (-1)^r}{2(n+r)}\left[S_2\left(\frac{n+r-1}{2}\right)
-\zeta(2)\right]\bigg{\}}$\\
&$+\frac{1}{2(n+r)^3}-\frac{\zeta(2)}{2(n+r)}$\\
&  \\[-3mm]
\hline
& \\[-3mm]
$ 2\ln(1+z){\rm Li}_2(-z)$ & 
$\frac{1}{n}\Bigg[\eta(n)\bigg{\{}-\zeta(2)S_{1}(n)-2S_{1}(n)\Big[
\frac{1+\eta(n)}{4}S_{2}\left(\frac{n}{2}\right)$\\
$+\ln z \ln^2(1+z)+2S_{1,2}(-z)$ &$+\frac{1-\eta(n)}{4}S_{2}\left(\frac{n-1}{2}\right)
-S_{2}(n)\Big]$\\
&$-\frac{1+\eta(n)}{8}S_{3}\left(\frac{n}{2}
\right)-\frac{1-\eta(n)}{8}S_{3}\left(\frac{n-1}{2}\right)$\\
&$+S_{3}(n)+\frac{2}{\eta^2(n)}\left(\widetilde{S}(n)+\frac{5\zeta(3)}{8}\right)
-\frac{3\zeta(3)}{4}\bigg{\}}$\\
&$-\bigg{\{}
-\eta(n)\left[S_3(n)+\zeta(3)\right]$\\
&$+\frac{1+\eta(n)}{8}\left[S_3\left(\frac{n}{2}\right)-\zeta(3)\right]$\\
&$-\frac{1-\eta(n)}{8}\left[S_3\left(\frac{n-1}{2}\right)-\zeta(3)\right]\bigg{\}}$\\
&$+\zeta(2)\bigg{\{}\eta(n)\left[\ln 2-S_1 (n)\right]
+\frac{1+\eta(n)}{2}S_1\left(\frac{n}{2}\right)$\\
&$-\frac{1-\eta(n)}{2}S_1\left(\frac{n-1}{2}\right)
\bigg{\}}
-\zeta(2)\ln 2+\frac{\zeta(3)}{4}\Bigg]$\\
& \\
\hline
\hline
\end{tabular}
\end{center}
\end{table}


\begin{thebibliography}{0}
\bibitem{footnote1}
Around the time that these
particle separated FFs were determined, FFs for 
charged hadrons were presented in Ref.\ \cite{Bourhis:2000gs}.
\bibitem{footnote3}
In Ref.\ \cite{Kniehl:2000fe}, a conceptual error was made. 
The parameters $\alpha$ and $\beta$ of $D_u^{p/\overline{p}}$ were also 
fixed to be equal to twice those of $D_d^{p/\overline{p}}$. This does not invalidate that fit,
but implies that the parameterizations for the proton FFs in Ref.\ \cite{Kniehl:2000hk} 
are valid only if this erroneous identification is made, which was not explicitly mentioned.
\bibitem{footnote4}
In \cite{MH}, the quantity 
$1/(2\pi N_{\rm events} p_T) (d^2 N/(dp_T dy))|_{y=0}$ 
is measured. These data must be multiplied by the measured non-singly diffractive
cross section of this experiment of $\sigma=30$ mb \cite{MH2}
to obtain the corresponding data for $E d^3 \sigma/dp^3$.
\end{thebibliography}
\end{document}